\documentclass[aps,reprint,superscriptaddress]{revtex4-1}
\pdfoutput=1

\usepackage{graphicx}
\usepackage{setspace}
\usepackage{lineno}
\usepackage{physics}
\usepackage{braket}

\makeatletter
\renewcommand{\fnum@figure}{\textbf{Fig.\thefigure}}
\makeatother

\begin{document}

\title{Quantum sensing of a coherent single spin excitation in a nuclear ensemble} 

\author{D. M. Jackson \textsuperscript{1*}}
\author{D. A. Gangloff \textsuperscript{1*$\dagger$}}
\author{J. H. Bodey \textsuperscript{1}}
\author{L. Zaporski \textsuperscript{1}}
\author{C. Bachorz \textsuperscript{1}}
\author{E. Clarke \textsuperscript{2}}
\author{M. Hugues \textsuperscript{3}}
\author{C. Le Gall \textsuperscript{1}}
\author{M. Atat\"ure\textsuperscript{1 $\dagger$}}

\noaffiliation

\affiliation{Cavendish Laboratory, University of Cambridge, JJ Thomson Avenue, Cambridge, CB3 0HE, UK}
\affiliation{EPSRC National Epitaxy Facility, University of Sheffield, Sheffield, Broad Lane, S3 7HQ, UK}
\affiliation{Universit\'e C\^ote d'Azur, CNRS, CRHEA, rue Bernard Gregory, 06560 Valbonne, France
\\ \ \\
\textsuperscript{*} These authors contributed equally to this work.
\\
\textsuperscript{$\dagger$} e-mail: dag50@cam.ac.uk; ma424@cam.ac.uk.
\\ \ \\
}

\begin{abstract}
The measurement of single quanta in a collection of coherently interacting objects is transformative in the investigations of emergent quantum phenomena. An isolated nuclear-spin ensemble is a remarkable platform owing to its coherence, but detecting its single spin excitations has remained elusive. Here, we use an electron spin qubit in a semiconductor quantum dot to sense a single nuclear-spin excitation (a nuclear magnon) with 1.9-ppm precision via the 200-kHz hyperfine shift on the 28-GHz qubit frequency. We demonstrate this single-magnon precision across multiple modes identified by nuclear species and polarity. Finally, we monitor the coherent dynamics of a nuclear magnon and the emergence of quantum correlations competing against decoherence. A direct extension of this work is to probe engineered quantum states of the ensemble including long-lived memory states. 
\end{abstract}

\maketitle



An ensemble of coherently interacting spins can host spectacular examples of spontaneous and driven quantum many-body phases, such as superradiance \cite{Dicke1954,Kaluzny1983} and time crystals \cite{Choi2017,Zhu2019}. Further, if one can access coherently the quanta of such ensembles they can be exploited for quantum computational tasks \cite{Rotondo2015,Taylor2003} and for the storage of quantum information in collective modes of the ensemble\textemdash a quantum memory \cite{Taylor2003b,Denning2019,Chaneliere2005,Tanji2009}. To this end a promising avenue is to leverage the remarkable coherence of a nuclear spin system \cite{Steger2012,Saeedi2013}, interfaced with an electron proxy qubit that allows for fast control and readout. Most notable progress has been made in the limit of countable nuclear spins coupled to a central electron; namely in colour centres within diamond and silicon carbide \cite{GurudevDutt2007,Taminiau2012,Taminiau2014,Metsch2019,Bourassa2020}, silicon-based quantum dots \cite{Hensen2020}, rare-earth defect spins \cite{Car2018} and donors in silicon \cite{Morton2008,Pla2013}. The semiconductor quantum dot (QD) is a natural realization of an altogether more dense, coherent nuclear ensemble \cite{Chekhovich2013,Chekhovich2015,Waeber2019,Wust2016}, again interfaced to a central-electron proxy qubit \cite{Stockill2016,Bechtold2015,Wust2016,Bethke2019,Greilich2007,Reilly2008,Vink2009,Bluhm2011}. The promise of this system lies in the recent demonstration of the deterministic injection of a single spin excitation -- a nuclear magnon -- into the ensemble \cite{Gangloffeaaw2906}. Complementing this control with the ability to detect, or sense, single quantum excitations can reveal directly the coherent internal dynamics of such a dense spin ensemble.

Quantum sensors, in their own right, have seen impressive leaps in the detection of few distinguishable nuclear spins using colour centres in diamond and silicon carbide \cite{Childress2006,Zhao2012,Kolkowitz2012,Shi2015,Lovchinsky2016,Nagy2019} and rare-earth defect spins \cite{Kornher2020}. In dense systems of indistinguishable spins, single-spin detection has only recently been achieved for electronic ensembles \cite{Lachance-Quirion2017,Lachance-Quirion2020}; in contrast, resolution for nuclear ensembles remains at the level of a few-hundred spins \cite{Probst2020}. The ability to sense a single spin excitation in a dense ensemble of indistinguishable nuclei, as desired to observe coherent many-body phenomena, has been an outstanding challenge.

\begin{figure*}
\includegraphics[width = 0.96\textwidth]{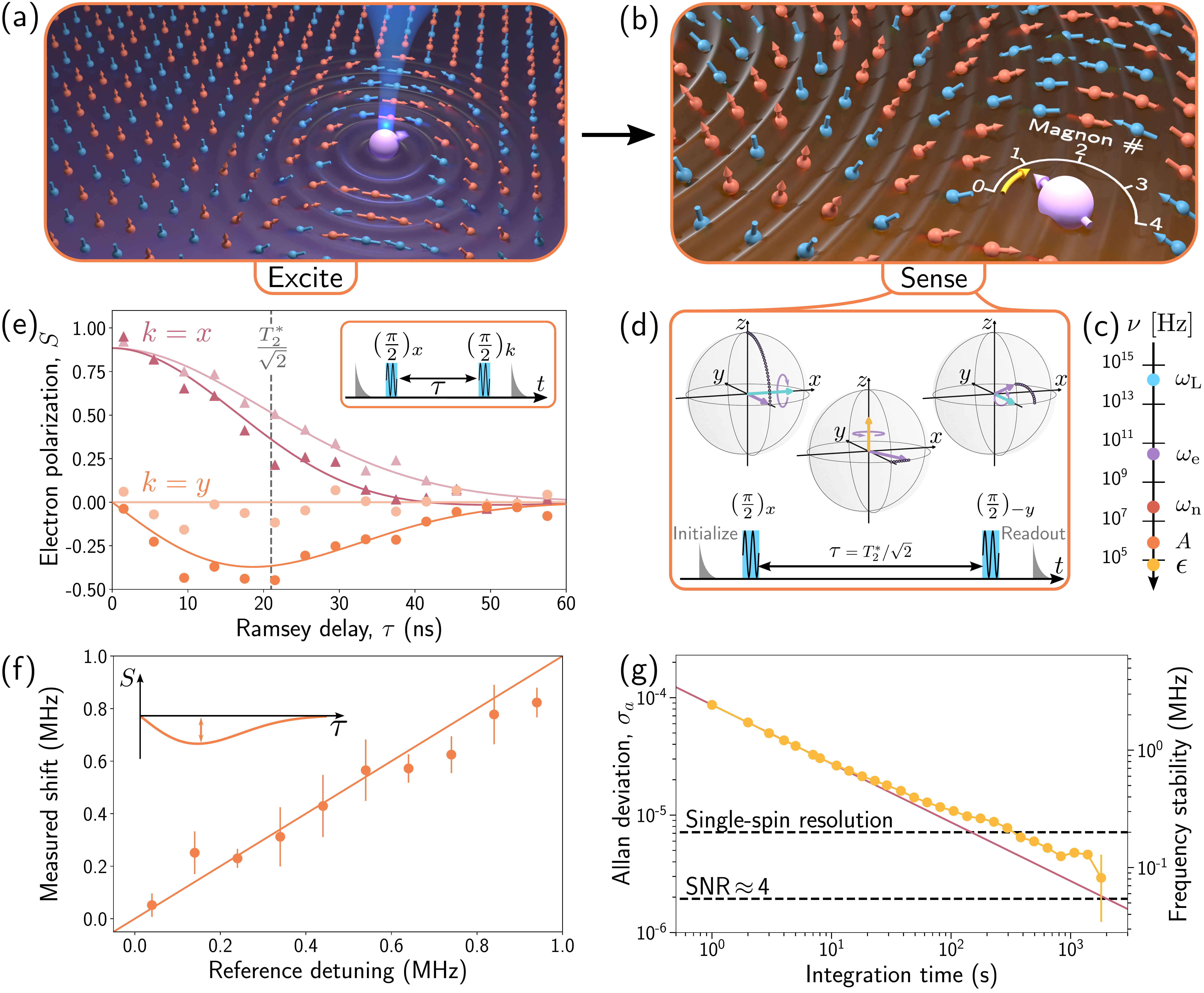}
\caption{\textbf{Quantum sensing with single-magnon precision.} \textbf{(a)} Central spin system of a proxy qubit coupled to a nuclear ensemble. Driving optically the proxy qubit injects a nuclear magnon into the ensemble. \textbf{(b)} Proxy qubit is used as a quantum sensor to detect nuclear magnons. \textbf{(c)}  Frequency hierarchy: laser $\omega_\text{L} = 315$\,THz, electron Zeeman $\omega_\text{e}=28$\,GHz, nuclear Zeeman $\omega_\text{n}\approx 35$\,MHz, hyperfine constant $A\approx 500$\,kHz, ESR-frequency precision $\epsilon=50$\,kHz. \textbf{(d)} \textit{Side-of-fringe} Ramsey interferometry. Readout/initialization is a resonant pulse from the spin-$\downarrow$ state to an optically excited state \cite{Atature2006}. With (without) a preceding $\pi$-pulse (not shown) this yields the electron spin population, $\rho_{\uparrow\uparrow}$ ($\rho_{\downarrow\downarrow}$), and polarization $S=\rho_{\downarrow\downarrow}-\rho_{\uparrow\uparrow}$ \cite{SI}. Bloch spheres: (left) a $(\frac{\pi}{2})_x$ rotation on a spin-$\uparrow$ electron creates a state along $-y$. (middle) The electron state precesses at a frequency $\Delta\omega$ around the effective magnetic field of a magnon (yellow arrow), accruing a phase $2\pi\Delta\omega\tau$. (right) $(\frac{\pi}{2})_{-y}$ rotation transfers this accrued phase to a $z$-basis polarization, $S$. \textbf{(e)} Electron polarization, $S$, versus $\tau$ for \textit{side-of-fringe} and \textit{top-of-fringe} Ramsey sequences (orange and pink, respectively) in the presence of a $0$\,MHz (light) and $-6$\,MHz (dark) ESR shift. Inset represents the pulse sequence for \textit{top-of-fringe} (\textit{side-of-fringe}) Ramsey for $k=x$ ($k=y$). Solid curves are theoretical fits with $T_2^* \approx 30$\,ns \cite{SI}. Grey line indicates the optimum delay, $T_2^*/\sqrt{2}$. \textbf{(f)} Ramsey-measured shift versus reference detuning from the electron rotating frame (circles). Solid curve is the one-to-one correspondence. \textbf{(g)} Allan deviation, $\sigma_\text{a}$, of Ramsey-measured shift (yellow circles). The pink solid curve is the quantum limit (photon shot noise). Statistical error bars represent one standard deviation (e, f).}
\label{figure1} 
\end{figure*}

Here, we achieve quantum sensing of a single nuclear-spin excitation in a dense ensemble of $\sim$80,000 nuclei. After injecting deterministically a single nuclear magnon into the cooled spin ensemble of a QD \cite{Gangloffeaaw2906} (Fig. \ref{figure1}a), we perform a Ramsey \textit{side-of-fringe} measurement of the electron spin resonance (ESR) frequency to sense this excitation (Fig. \ref{figure1}b). Our resolution of $50$\,kHz on a $28$-GHz electron spin splitting, or $1.9$ parts per million (ppm), enables us to sense the $200$-kHz hyperfine shift induced by a single nuclear magnon. Furthermore, we resolve multiple magnon modes distinguished by atomic species and spin polarity via the spectral dependence of this hyperfine shift. Finally, we observe the time-dependent shift induced on our sensor by collective Rabi oscillations, as the electron qubit coherently brings a nuclear magnon in and out of existence. These dynamics reveal the competition between the buildup of quantum correlations and decoherence in this dense spin ensemble.


Our system consists of an optically active central electron spin coupled via the hyperfine interaction to $N$\,$\sim$\,$80,000$ nuclear spins of In (total spin $I=9/2$), Ga ($I=3/2$), and As ($I=3/2$). The nuclear ensemble is prepared optically, via the electron, in a reduced fluctuation state at zero polarization \cite{Ethier-Majcher2017,Gangloffeaaw2906}. The Zeeman-split electron spin is driven with microwave-modulated optical fields for fast, multi-axis control \cite{Bodey2019}. All spin operations occur in a frame rotating at the effective microwave frequency, chosen to match the ESR. The electron qubit then generates nuclear spin excitations thanks to an activated noncollinear hyperfine interaction, arising from nuclear quadrupolar effects and strain \cite{Hogele2012,Gangloffeaaw2906}. An approximate yet elegant Hamiltonian gives an intuitive picture of the terms relevant to control and sensing of the nuclear spin system \cite{SI}, as expressed in the rotating frame of the electron spin with $z$ as its quantization axis:
\begin{equation}
	H=\underbrace{\delta S_z + \Omega S_x}_{\text{qubit\ drive}} - \underbrace{\eta\Omega S_y M^{\pm}}_{\substack{\text{activated} \\ \text{exchange}}} +\underbrace{\omega_\text{n} I_z}_{\substack{\text{nuclear} \\ \text{Zeeman}}} - \underbrace{2A S_z I_z}_{\text{sensing}}. 
\label{equation_system_hamiltonian}
\end{equation}
\noindent
$S_i$ are the electron-spin Pauli operators and $I_z$ is the component of the total nuclear spin along the external magnetic field; $\delta$ and $\Omega$ are the detuning and Rabi frequency of our \textit{qubit drive}. $M^{\pm}$ dictates single-magnon creation and annihilation that change $I_z$ by one unit, exchanged against an electronic spin-flip activated by the \textit{qubit drive} with a rate $\eta \Omega$. The dimensionless constant $\eta< 1$ captures the strength of the noncollinear hyperfine interaction \cite{Gangloffeaaw2906,SI}. This \textit{activated exchange} is resonant at the Hartmann-Hahn condition $\Omega^2 + \delta^2 = \omega_\text{n}^2$ \cite{Hartmann1962}, where the driven qubit's generalized Rabi frequency matches the \textit{nuclear Zeeman} energy ($\omega_\text{n}$) and thereby supplies the energy required for a single nuclear-spin flip \cite{Gangloffeaaw2906,Bodey2019}. The \textit{sensing} term of Eq. \ref{equation_system_hamiltonian} couples the electron to an effective magnetic field $2A I_z$ through the hyperfine interaction; $A$ is the single-nucleus hyperfine constant. Accordingly, the ESR frequency acts as a magnetic-field sensor, and we measure the change in $I_z$ of one unit induced by a single nuclear-spin excitation. Any such change is superimposed on an applied external field of $4.5$\,T, which sets the electronic Zeeman frequency, $\omega_\text{e}$, at $28$\,GHz. In our system, the ESR shift we predict from a single magnon is $\sim$$200$\,kHz \cite{Urbaszek2013}, underscoring the necessity for a sensing resolution in the ppm regime, as shown in the frequency hierarchy of Fig. \ref{figure1}c.

To measure the ESR-frequency shift with sub-$200$\,kHz precision we employ two-axis Ramsey interferometry in the maximally sensitive \textit{side-of-fringe} configuration \cite{Degen2017} (Fig. 1d, Fig. \ref{figure1}e circles). Within the available spin coherence time, $T_2^* \approx 30$\,ns (Fig. \ref{figure1}e, triangles), a Ramsey delay of $\tau=T_2^*/\sqrt{2}$ maximizes sensitivity. This control sequence, illustrated in Fig. \ref{figure1}d, maps linearly a small frequency shift to a change in electronic polarization in the $z$-basis.  A simple test of this sensing capability is made in Fig. \ref{figure1}f, whereby we apply a reference detuning during the sequence. The resulting phase accumulation of the electron relative to the drive mimics a magnon-induced ESR shift. The correspondence between this reference detuning and the shift measured by our sensor confirms the linearity of this mapping over a megahertz range.

A robust metric to benchmark our sensor performance is the Allan deviation, which measures the variation in the electron spin-splitting from run-to-run as a function of integration time \cite{Allan1966,Degen2017}. 
As shown in Fig. \ref{figure1}g, the Allan deviation initially scales inversely with the square root of the integration time, as expected from the quantum-limited noise of our photonic readout. Beyond $20$\,s of integration, slow noise on the ESR frequency starts to dominate. At an integration time of $2000$\,s we reach a minimum detectable shift for a single measurement of the ESR frequency of $100$\,kHz, or twice the extrapolated quantum limit of $50$\,kHz. Finally, we combine a \textit{reference} and \textit{sense} measurement before and after the injection of a single magnon, $\Delta\omega_\text{ref}$ and $\Delta\omega_\text{sense}$ respectively, and take their difference $\Delta\omega_D=\Delta\omega_\text{sense}-\Delta\omega_\text{ref}$ on a microsecond timescale. This step brings us down to the quantum limit with a differential ESR frequency precision of $1.9$\,ppm, and a signal-to-noise ratio of $4$ for the anticipated single magnon shift of $200$\,kHz.

\begin{figure} 
\includegraphics[width = \columnwidth]{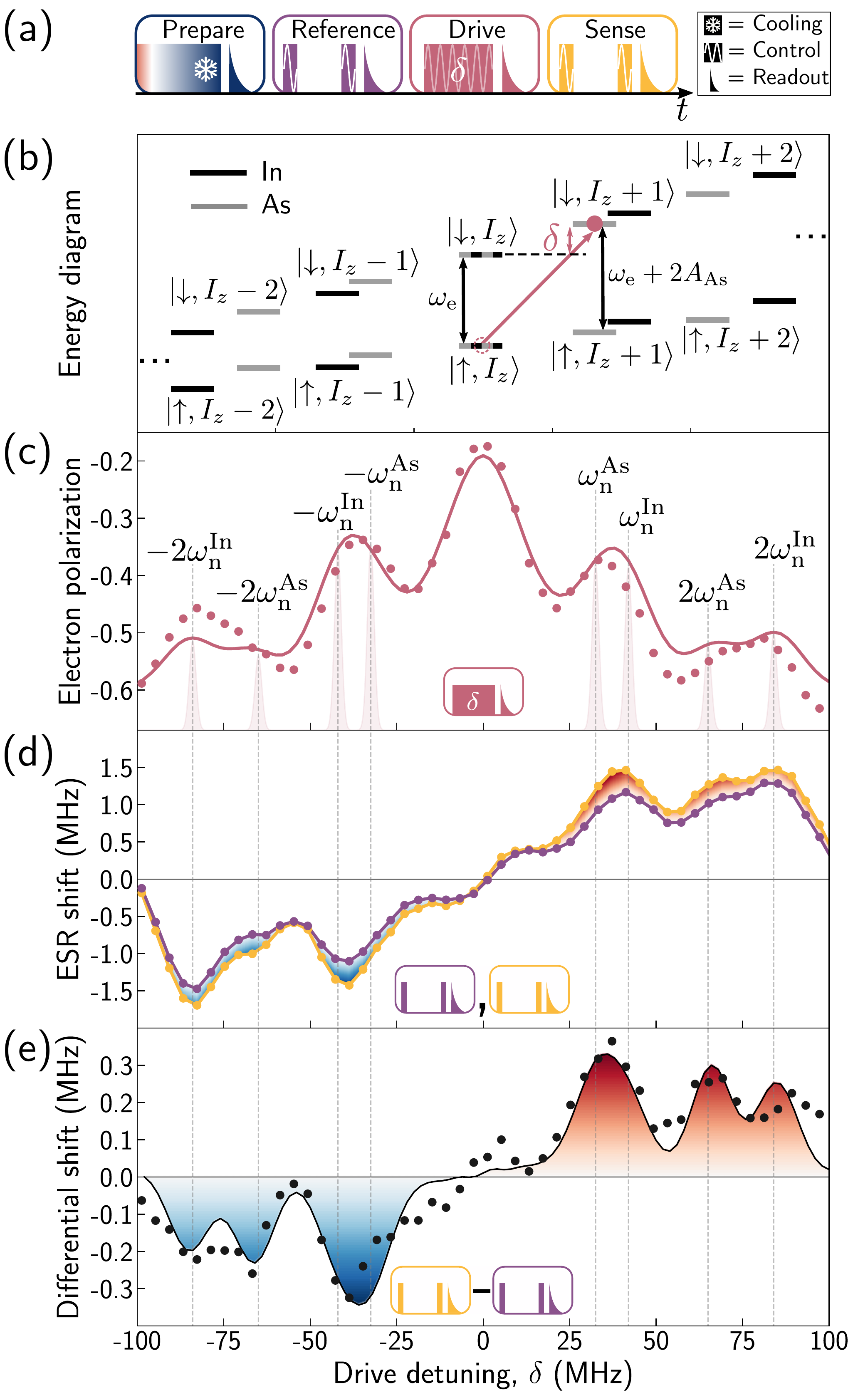}
\caption{\textbf{Single-magnon spectrum.} \textbf{(a)} Experimental pulse sequence: (1) $11\,\mu$s of optimal nuclear spin cooling \cite{Gangloffeaaw2906,SI}, (2) \textit{reference} Ramsey measurement, (3) $1.8\,\mu$s of \textit{qubit drive} with detuning $\delta$ and fixed $\Omega = 8$\,MHz, followed by a readout, (4) \textit{sense} Ramsey measurement. \textbf{(b)} Lab-frame eigenstates of the undriven electron spin dressed with collective nuclear states, $I_z$, of arsenic (grey) and indium (black). Pink arrow illustrates a drive pulse with detuning $\delta$ activating a change in $I_z$. Thanks to the hyperfine interaction, $-2A I_z S_z$, the electron spin splitting changes. \textbf{(c)} Electron polarization, $S$, measured directly after drive (circles). Solid curve is our master equation model \cite{SI}; sidebands occur at once or twice the nuclear Zeeman frequencies ($\omega_\text{n}^{\text{As}}=32.5$\,MHz, $\omega_\text{n}^{\text{In}}=42$\,MHz), illustrated by the shaded peaks below. \textbf{(d)} \textit{Reference} (purple circles) and \textit{sense} (yellow circles) signals, $\Delta\omega_{\text{ref}}$ and $\Delta\omega_{\text{sense}}$ respectively. \textbf{(e)} The differential shift $\Delta\omega_{\text{D}}=\Delta\omega_{\text{sense}}-\Delta\omega_{\text{ref}}$ (black circles). Solid curve is our master equation model \cite{SI}. (c), (d) and (e) are measured as a function of detuning, $\delta$.}
\label{figure2}
\end{figure} 

\begin{figure*}
\includegraphics[width = \textwidth]{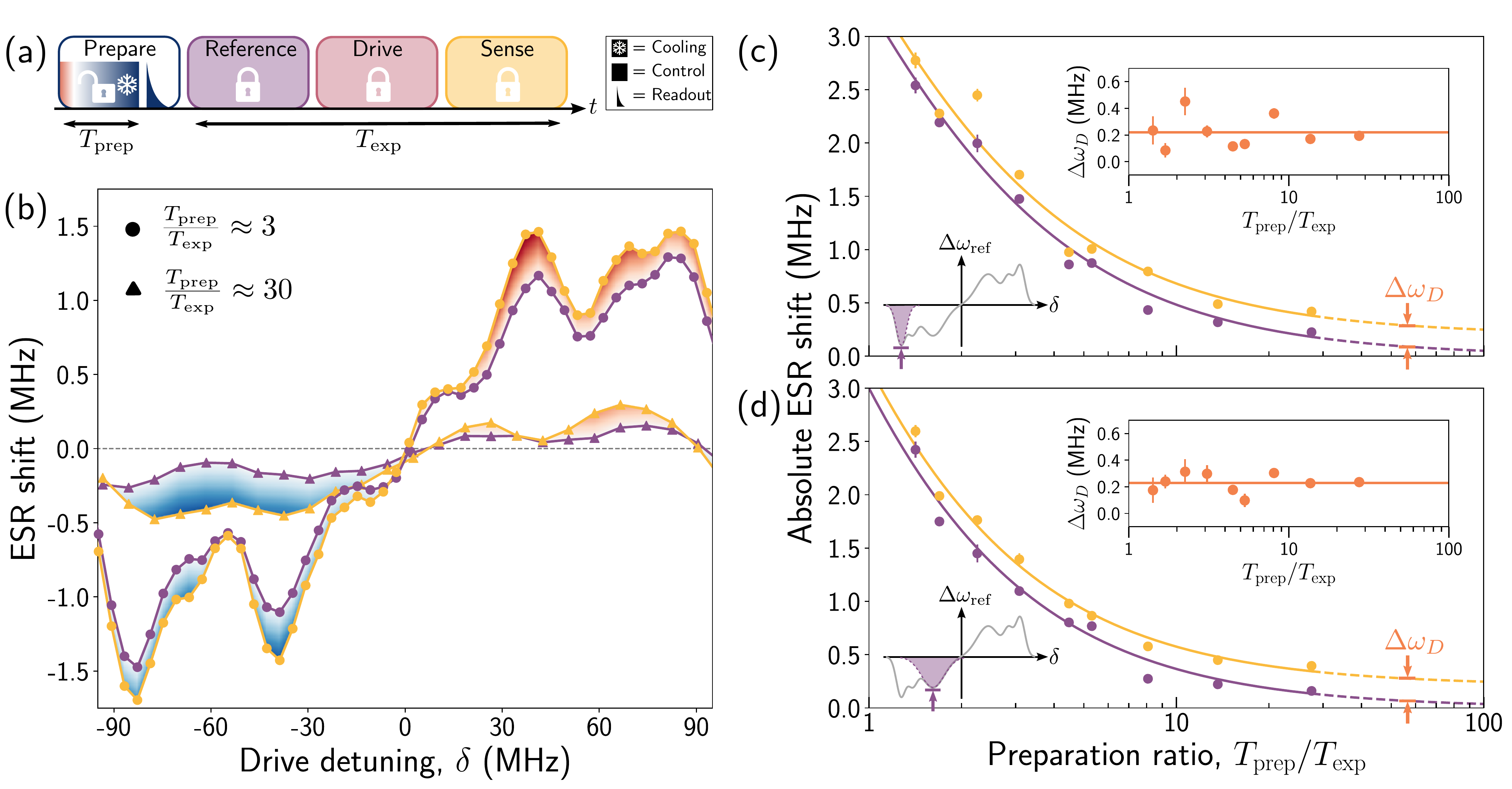}
\caption{\textbf{Erasing a single magnon.} \textbf{(a)} Preparation time $T_\text{prep}$ is tuned from $5.1\,\mu$s to $99.1\,\mu$s  within an otherwise fixed experimental sequence. The reference, drive, and sense steps are the same as for Fig. 2. \textbf{(b)} Detuning dependence of ESR shifts $\Delta\omega_\text{ref}$ (purple) and $\Delta\omega_\text{sense}$ (yellow) for two preparation ratios: $T_\text{prep}/T_\text{exp}\approx3$ and $T_\text{prep}/T_\text{exp}\approx30$ (circles and triangles, respectively). \textbf{(c),} \textbf{(d)} $\Delta\omega_\text{ref}$ (purple circles) and $\Delta\omega_\text{sense}$ (yellow circles) as a function of preparation ratio, for $I_z-2$ (c) and $I_z-1$ (d) magnon modes.  Solid curves are fits to a phenomenological saturation model \cite{SI}. Differential shift, $\Delta\omega_\text{D}$, is shown as a function of preparation ratio (inset); the data have zero slope within error and we show the mean as solid curves.}
\label{figure3}
\end{figure*}

With a fully characterized quantum sensor, we now introduce nuclear magnons to the picture with the \textit{activated exchange} term in the Hamiltonian (Eq. \ref{equation_system_hamiltonian}). In the experimental sequence of Fig. \ref{figure2}a, this comprises the inclusion of a magnon-injection pulse between the reference and sensing measurements. By setting the drive detuning, $\delta=\pm\omega_{\text{n}}, \pm2\omega_{\text{n}}$, we can address the $I_z\rightarrow I_z\pm1$ and $I_z\rightarrow I_z\pm2$ transitions selectively and resonantly, as seen in Fig. \ref{figure2}a -- the quadrupolar nature of the \textit{activated exchange} permits single excitations in both $\pm1$ and $\pm2$ magnon modes \cite{Gangloffeaaw2906}. Reading out the electron spin directly after the driving reveals the expected electron-polarization spectrum (Fig. \ref{figure2}c), evidencing magnonic injection sidebands at the nuclear Zeeman frequencies ($\omega_\text{n}^{\text{As}}=32.5$\,MHz, $\omega_\text{n}^{\text{In}}=42$\,MHz) \cite{Gangloffeaaw2906}. Following our optimized sensing prescription (Fig. \ref{figure1}), Fig. \ref{figure2}d presents the raw frequency shifts measured during the \textit{reference} and \textit{sense} sequences (purple and yellow, respectively). These signals include both slow noise and residual nuclear polarization; their difference, $\Delta\omega_D$, reveals exclusively the frequency shift due to a single nuclear magnon. 

Figure \ref{figure2}e presents this measured $\Delta\omega_D$ as a function of drive detuning, revealing single excitations in distinct magnon modes. We observe both positive and negative shifts of $\sim200$\,kHz, as expected from excitations injected by driving the positive and negative magnon sidebands, respectively. Indeed, population in the natural magnonic basis, proportional to $I_z$, maps directly to the electron-spin splitting via the hyperfine interaction, and the polarity of the shift makes positive and negative nuclear spin excitations distinguishable. Further, the power of our differential measurement technique is made doubly explicit, as the $50$\,kHz precision allows us to resolve clearly the lower contrast features making up the $I_z\rightarrow I_z\pm2$ peaks. These correspond to the $2\omega_\text{n}^{\text{As}}$ and $2\omega_\text{n}^{\text{In}}$ resonances of arsenic and indium, demonstrating the ability to excite and sense single collective spin excitations selectively within sub-ensembles of distinct nuclear species. 

Intuitively, the \textit{sense} signal (yellow circles, Fig. \ref{figure2}d) and the differential signal (Fig. \ref{figure2}e) exhibit magnon resonances; counter-intuitively, so does the \textit{reference} signal (purple circles, Fig. \ref{figure2}d). Ideally, the preparation step, which cools the nuclear spins, fully erases the magnon injected during the previous shot of the experiment, keeping $\Delta\omega_\text{ref}$ independent of $\delta$. In practice, owing to the finite preparation time, a residual nuclear polarization accumulates until a steady state that balances cooling and driving is reached -- an overall process equivalent to dynamic nuclear polarization \cite{Abragam1961}. As a result, a cumulative shift of $\sim1$\,MHz appears in both the reference and sensing measurements when $\delta$ meets a magnon resonance (Fig. \ref{figure2}d). With an asymptotically long preparation step, this cumulative shift disappears; its value depends on the time ratio of nuclear spin preparation to magnon drive, $T_\text{prep}/T_\text{exp}$ (Fig. \ref{figure3}a). Figure \ref{figure3}b shows the cumulative shift spectra for $T_\text{prep}/T_\text{exp}\approx3$ (circles) and $T_\text{prep}/T_\text{exp}\approx30$ (triangles); for the larger ratio the cumulative shift is eliminated almost entirely such that $\Delta\omega_\text{ref}$ is less than the single-magnon shift. Figures \ref{figure3}c and \ref{figure3}d summarize the complete dependence of $\Delta\omega_\text{ref}$ and $\Delta\omega_\text{sense}$ on the preparation ratio, for both $I_z-2$ and $I_z-1$ magnon modes, respectively. Extrapolating the fitted curves to high preparation ratio, we see that $\Delta\omega_\text{ref}$ ($\Delta\omega_\text{sense}$) approaches zero (the single-magnon shift). As a side note, this cumulative polarization at small preparation ratios can also be used to amplify the magnon signal. Importantly, the differential signal $\Delta\omega_\text{D} = \Delta\omega_\text{ref}-\Delta\omega_\text{sense}$ is independent of the preparation ratio, as demonstrated by the insets of Figs. \ref{figure3}c and 3d, verifying that $\Delta\omega_\text{D}$ measures the single magnon shift. This untainted detection of single magnons through the differential signal allows us in the experiments that follow to operate in a low preparation regime $T_\text{prep}/T_\text{exp}\approx3$, which is favourable for sensitivity.

A master-equation simulation of the electron and two nuclear species (As and In, as in Fig. \ref{figure2}b), in a truncated Hilbert space of $50$ states, is used to fit the single-magnon spectrum \cite{SI}. Taking into account spin decoherence and relaxation, which reduce the sensitivity of our measurements, we arrive at the species-specific single-nucleus hyperfine interaction constants for our magnons, $A_{\text{As}}=950$\,kHz and $A_{\text{In}}=550$\,kHz \cite{SI}. With these precise estimates, our quantum sensor is now further calibrated to convert an ESR shift to a magnonic population in the $\ket{I_z\pm k}$ excited state $\rho_{\pm k}=\pm\Delta\omega_\text{D}/2kA$.

\begin{figure} 
\includegraphics[width =\columnwidth]{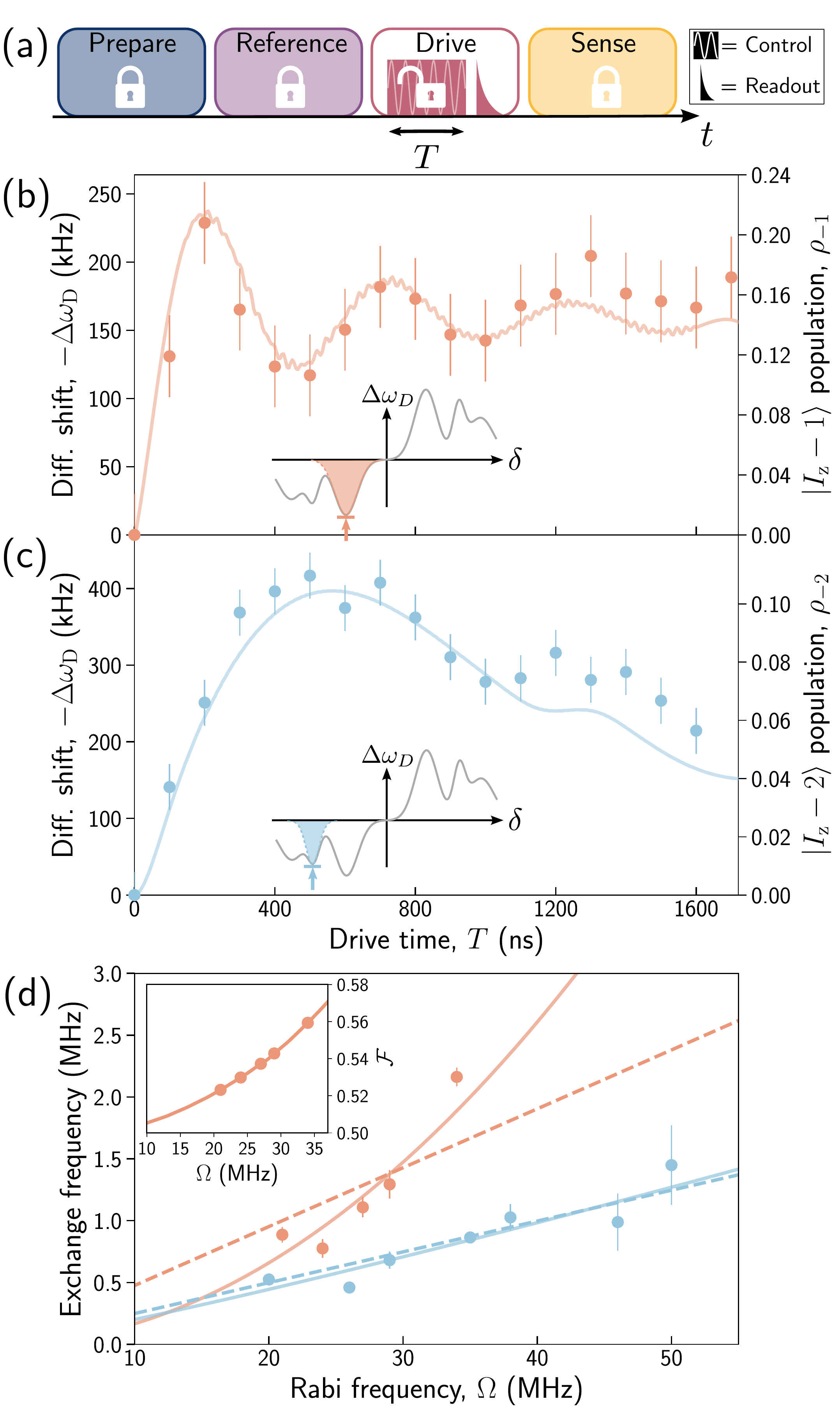}
\caption{\textbf{Sensing coherent dynamics of a dense nuclear ensemble.} \textbf{(a)} Drive time $T$ is tuned from $0\,\mu$s to $1.7\,\mu$s  within an otherwise fixed experimental sequence. The prepare, reference and sense steps are the same as for Fig. 2.  \textbf{(b),} \textbf{(c)} Circles present differential shift, $\Delta\omega_\text{D}$, with a 3-point running average, as a function of drive time $T$ for (b) $\Omega = 34$\,MHz and $\delta = 21$\,MHz, resonant with the $I_z-1$ mode at $\omega_\text{n}^\text{In} = 42$\,MHz; (c) $\Omega = 35$\,MHz and $\delta = 51$\,MHz, resonant with the $I_z-2$ mode at $2\omega_\text{n}^\text{As} = 65$\,MHz. The error bars are $\sigma_a$. Solid curves are results of our $50$-state master equation model. We use $A_{\text{In}}$ and $A_{\text{As}}$ from this model to convert $\Delta\omega_\text{D}$ to magnonic population $\rho_{-1}$ and $\rho_{-2}$ (right axis), respectively. \textbf{(d)} Activated exchange frequency versus qubit-drive Rabi frequency, $\Omega$, for $I_z-1$ (orange circles) and $I_z-2$ (blue circles) magnon modes. Error bars represent the $68\%$ confidence interval from the fitted Rabi oscillations \cite{SI}. Dashed curves are linear fits, $\eta\Omega$, with $\eta_1=4.8(6)\cdot10^{-2}$ (orange) and $\eta_2=2.5(1)\cdot10^{-2}$ (blue). Solid curves are superlinear best fits $\alpha\Omega^{\beta}$; $\alpha_1=0.002(^{+8}_{-1})$ and $\beta_1=2.0(5)$ (orange); $\alpha_2=0.014(^{+17}_{-8})$ and $\beta_2=1.1(2)$ (blue). Inset shows state overlap $\mathcal{F}$ as a function of $\Omega$.} 
\label{figure4}
\end{figure}

Figure 4 presents an elementary benchmark for sensing coherent dynamics of a dense nuclear ensemble: Rabi oscillations of a single magnon. Via the \textit{activated exchange} term of Eq. \ref{equation_system_hamiltonian} we drive a magnon mode resonantly for a time $T$, before sensing the resulting magnon population, $\rho_{\pm k}$ (Fig. \ref{figure4}a). In Figs. \ref{figure4}b and \ref{figure4}c, we observe coherent oscillations in the differential shift -- and equivalently magnon population -- as we drive Rabi flopping within the indium $\ket{I_z}\leftrightarrow \ket{I_z-1}$ and arsenic $\ket{I_z}\leftrightarrow \ket{I_z-2}$ modes, respectively. From Eq. 1, the \textit{activated exchange} frequency is simply $\eta\Omega$ \cite{SI}, where $\eta \propto \sqrt{N}$ captures the relative amplitude of a single nuclear-spin flip that is collectively enhanced by the presence of $N$ spins in the same mode \cite{Gangloffeaaw2906}. Figure \ref{figure4}d shows the $\Omega$-dependence of the \textit{activated exchange} frequency for both modes, with $\eta_1=4.8(6)\cdot10^{-2}$ and $\eta_2=2.5(1)\cdot10^{-2}$, presuming a linear dependence. This is equivalent to an exchange between the electron and a nuclear spin excitation distributed over an effective number $N$\,$\sim$\,$300$ and $N$\,$\sim$\,$200$ fully coherent indium and arsenic nuclei, respectively. That said, a closer inspection of Fig. \ref{figure4}D reveals that the \textit{activated exchange} frequency deviates from the presumed linear $\Omega$ dependence and fits best to $\Omega^{2.0}$ ($\Omega^{1.1}$) for the $I_z-1$ ($I_z-2$) mode. Such a nonlinearity could arise if the emerging collective mode competes against ensemble inhomogeneities \cite{Heidemann2007}, and the size of our magnon mode, $N$, grows with the drive strength, $\Omega$ \cite{SI}. Consequently, this observed superlinear Rabi-frequency scaling $\Omega\sqrt{N(\Omega)} = O(\Omega^{\beta}), \beta>1$ suggests the presence of drive-stimulated collective phenomena beyond magnon formation.

This drive-enhanced magnon coherence in turn improves the degree of entanglement of the proxy qubit with the nuclear ensemble. We quantify this via the state overlap, $\mathcal{F}=\bra{\psi} \rho \ket{\psi}_{\theta, \phi}$ \cite{Nielsen2000}, of a generalized electron-nuclear Bell state $\ket{\psi} = \exp(i\theta S_x/2)\left[\ket{\downarrow,I_z} + \exp(i\phi)\ket{\uparrow,I_z-1}\right]$ with the system's full density matrix, $\rho$, extracted from fits to Figs. \ref{figure4}b and \ref{figure4}c (solid curves) \cite{SI}; $\theta$ represents a rotation of the electron spin and $\phi$ is a relative phase. The inset of Fig. \ref{figure4}d shows that $\mathcal{F}$ beats the non-classicality threshold of $0.5$ for each of our drive strengths (orange circles), and grows with drive strength, reaching $0.56$ at our maximum Rabi frequency $\Omega=34$ MHz and $T=100$\,ns $\sim 1/4\eta\Omega$. This places the magnon modes of our system comfortably in the quantum-correlated regime.

With proxy-qubit sensing of single spin excitations in a dense nuclear ensemble we have opened a high-precision window into the many-body dynamics of this strongly interacting system. Our results already reveal species-selective coherent dynamics and the emergence of drive-enhanced quantum correlations. In the future the same tools can be applied to the investigation of spin-wave superposition states and the verification of quantum-state transfer into long-lived collective memory states. With the addition of global nuclear-spin control, this work can also be expanded towards fully fledged tomography of strongly correlated many-body states.


\bibliographystyle{science}
\bibliography{ms}

\begin{thebibliography}{10}

\bibitem{Dicke1954}
R.~H. Dicke, {\it Phys. Rev.\/} {\bf 93}, 99 (1954).

\bibitem{Kaluzny1983}
Y.~Kaluzny, P.~Goy, M.~Gross, J.~M. Raimond, S.~Haroche, {\it Phys. Rev.
  Lett.\/} {\bf 51}, 1175 (1983).

\bibitem{Choi2017}
S.~Choi, {\it et~al.\/}, {\it Nature\/} {\bf 543}, 221 (2017).

\bibitem{Zhu2019}
B.~Zhu, J.~Marino, N.~Y. Yao, M.~D. Lukin, E.~A. Demler, {\it New J. Phys.\/}
  {\bf 21}, 073028 (2019).

\bibitem{Rotondo2015}
P.~Rotondo, M.~{Cosentino Lagomarsino}, G.~Viola, {\it Phys. Rev. Lett.\/} {\bf
  114}, 143601 (2015).

\bibitem{Taylor2003}
J.~M. Taylor, A.~Imamoglu, M.~D. Lukin, {\it Phys. Rev. Lett.\/} {\bf 91},
  246802 (2003).

\bibitem{Taylor2003b}
J.~M. Taylor, C.~M. Marcus, M.~D. Lukin, {\it Phys. Rev. Lett.\/} {\bf 90},
  206803 (2003).

\bibitem{Denning2019}
E.~V. Denning, D.~A. Gangloff, M.~Atat{\"{u}}re, J.~M{\o}rk, C.~{Le Gall}, {\it
  Phys. Rev. Lett.\/} {\bf 123}, 140502 (2019).

\bibitem{Chaneliere2005}
T.~Chaneli{\`{e}}re, {\it et~al.\/}, {\it Nature\/} {\bf 438}, 833 (2005).

\bibitem{Tanji2009}
H.~Tanji, S.~Ghosh, J.~Simon, B.~Bloom, V.~Vuleti{\'{c}}, {\it Phys. Rev.
  Lett.\/} {\bf 103}, 043601 (2009).

\bibitem{Steger2012}
M.~Steger, {\it et~al.\/}, {\it Science\/} {\bf 336}, 1280 (2012).

\bibitem{Saeedi2013}
K.~Saeedi, {\it et~al.\/}, {\it Science\/} {\bf 342}, 830 (2013).

\bibitem{GurudevDutt2007}
M.~V. {Gurudev Dutt}, {\it et~al.\/}, {\it Science\/} {\bf 316}, 1312 (2007).

\bibitem{Taminiau2012}
T.~H. Taminiau, {\it et~al.\/}, {\it Phys. Rev. Lett.\/} {\bf 109}, 137602
  (2012).

\bibitem{Taminiau2014}
T.~H. Taminiau, J.~Cramer, T.~{Van Der Sar}, V.~V. Dobrovitski, R.~Hanson, {\it
  Nat. Nanotechnol.\/} {\bf 9}, 171 (2014).

\bibitem{Metsch2019}
M.~H. Metsch, {\it et~al.\/}, {\it Phys. Rev. Lett.\/} {\bf 122}, 190503
  (2019).

\bibitem{Bourassa2020}
A.~Bourassa, {\it et~al.\/}, {\it arXiv\/} {\bf 2005.07602} (2020).

\bibitem{Hensen2020}
B.~Hensen, {\it et~al.\/}, {\it Nat. Nanotechnol.\/} {\bf 15}, 13 (2020).

\bibitem{Car2018}
B.~Car, L.~Veissier, A.~Louchet-Chauvet, J.~L. {Le Gou{\"{e}}t},
  T.~Chaneli{\`{e}}re, {\it Phys. Rev. Lett.\/} {\bf 120}, 197401 (2018).

\bibitem{Morton2008}
J.~J. Morton, {\it et~al.\/}, {\it Nature\/} {\bf 455}, 1085 (2008).

\bibitem{Pla2013}
J.~J. Pla, {\it et~al.\/}, {\it Nature\/} {\bf 496}, 334 (2013).

\bibitem{Chekhovich2013}
E.~A. Chekhovich, {\it et~al.\/}, {\it Nat. Mater.\/} {\bf 12}, 494 (2013).

\bibitem{Chekhovich2015}
E.~A. Chekhovich, M.~Hopkinson, M.~S. Skolnick, A.~I. Tartakovskii, {\it Nat.
  Commun.\/} {\bf 6}, 6348 (2015).

\bibitem{Waeber2019}
A.~M. Waeber, {\it et~al.\/}, {\it Nat. Commun.\/} {\bf 10}, 3157 (2019).

\bibitem{Wust2016}
G.~W{\"{u}}st, {\it et~al.\/}, {\it Nat. Nanotechnol.\/} {\bf 11}, 885 (2016).

\bibitem{Stockill2016}
R.~Stockill, {\it et~al.\/}, {\it Nat. Commun.\/} {\bf 7}, 12745 (2016).

\bibitem{Bechtold2015}
A.~Bechtold, {\it et~al.\/}, {\it Nat. Phys.\/} {\bf 11}, 1005 (2015).

\bibitem{Bethke2019}
P.~Bethke, {\it et~al.\/}, {\it Phys. Rev. Lett.\/} {\bf 125}, 047701 (2020).

\bibitem{Greilich2007}
A.~Greilich, {\it et~al.\/}, {\it Science\/} {\bf 317}, 1896 (2007).

\bibitem{Reilly2008}
D.~J. Reilly, {\it et~al.\/}, {\it Science\/} {\bf 321}, 817 (2008).

\bibitem{Vink2009}
I.~T. Vink, {\it et~al.\/}, {\it Nat. Phys.\/} {\bf 5}, 764 (2009).

\bibitem{Bluhm2011}
H.~Bluhm, {\it et~al.\/}, {\it Nat. Phys.\/} {\bf 7}, 109 (2011).

\bibitem{Gangloffeaaw2906}
D.~A. Gangloff, {\it et~al.\/}, {\it Science\/} {\bf 364}, 62 (2019).

\bibitem{Childress2006}
L.~Childress, {\it et~al.\/}, {\it Science\/} {\bf 314}, 281 (2006).

\bibitem{Zhao2012}
N.~Zhao, {\it et~al.\/}, {\it Nat. Nanotechnol.\/} {\bf 7}, 657 (2012).

\bibitem{Kolkowitz2012}
S.~Kolkowitz, Q.~P. Unterreithmeier, S.~D. Bennett, M.~D. Lukin, {\it Phys.
  Rev. Lett.\/} {\bf 109}, 137601 (2012).

\bibitem{Shi2015}
F.~Shi, {\it et~al.\/}, {\it Science\/} {\bf 347}, 1135 (2015).

\bibitem{Lovchinsky2016}
I.~Lovchinsky, {\it et~al.\/}, {\it Science\/} {\bf 351}, 836 (2016).

\bibitem{Nagy2019}
R.~Nagy, {\it et~al.\/}, {\it Nat. Commun.\/} {\bf 10}, 1954 (2019).

\bibitem{Kornher2020}
T.~Kornher, {\it et~al.\/}, {\it Phys. Rev. Lett.\/} {\bf 124}, 170402 (2020).

\bibitem{Lachance-Quirion2017}
D.~Lachance-Quirion, {\it et~al.\/}, {\it Sci. Adv.\/} {\bf 3}, e1603150
  (2017).

\bibitem{Lachance-Quirion2020}
D.~Lachance-Quirion, {\it et~al.\/}, {\it Science\/} {\bf 367}, 425 (2020).

\bibitem{Probst2020}
S.~Probst, {\it et~al.\/}, {\it arXiv\/} {\bf 2001.04854} (2020).

\bibitem{Atature2006}
M.~Atat{\"{u}}re, {\it et~al.\/}, {\it Science\/} {\bf 312}, 551 (2006).

\bibitem{SI}
Supplementary information.

\bibitem{Ethier-Majcher2017}
G.~{\'{E}}thier-Majcher, {\it et~al.\/}, {\it Phys. Rev. Lett.\/} {\bf 119},
  130503 (2017).

\bibitem{Bodey2019}
J.~H. Bodey, {\it et~al.\/}, {\it npj Quantum Inf.\/} {\bf 5}, 95 (2019).

\bibitem{Hogele2012}
A.~H{\"{o}}gele, {\it et~al.\/}, {\it Phys. Rev. Lett.\/} {\bf 108}, 197403
  (2012).

\bibitem{Hartmann1962}
S.~R. Hartmann, E.~L. Hahn, {\it Phys. Rev.\/} {\bf 128}, 2042 (1962).

\bibitem{Urbaszek2013}
B.~Urbaszek, {\it et~al.\/}, {\it Rev. Mod. Phys.\/} {\bf 85}, 79 (2013).

\bibitem{Degen2017}
C.~L. Degen, F.~Reinhard, P.~Cappellaro, {\it Rev. Mod. Phys.\/} {\bf 89},
  035002 (2017).

\bibitem{Allan1966}
D.~W. Allan, {\it Proc. IEEE\/} {\bf 54}, 221 (1966).

\bibitem{Abragam1961}
A.~Abragam, L.~C. Hebel, {\it Am. J. Phys.\/} {\bf 29}, 860 (1961).

\bibitem{Heidemann2007}
R.~Heidemann, {\it et~al.\/}, {\it Phys. Rev. Lett.\/} {\bf 99}, 163601 (2007).

\bibitem{Nielsen2000}
M.~A. Nielsen, I.~L. Chuang, {\it {Quantum Computation and Quantum
  Information}\/} (Cambridge University Press, 2000).

\end{thebibliography}


\begin{thebibliography}{10}

\bibitem{Stockill2016}
R.~Stockill, {\it et~al.\/}, {\it Nat. Commun.\/} {\bf 7}, 12745 (2016).

\bibitem{Ethier-Majcher2017}
G.~{\'{E}}thier-Majcher, {\it et~al.\/}, {\it Phys. Rev. Lett.\/} {\bf 119},
  130503 (2017).

\bibitem{Gangloff2019}
D.~A. Gangloff, {\it et~al.\/}, {\it Science\/} {\bf 364}, 62 (2019).

\bibitem{Urbaszek2013}
B.~Urbaszek, {\it et~al.\/}, {\it Rev. Mod. Phys.\/} {\bf 85}, 79 (2013).

\bibitem{Degen2017}
C.~L. Degen, F.~Reinhard, P.~Cappellaro, {\it Rev. Mod. Phys.\/} {\bf 89},
  035002 (2017).

\bibitem{Allan1966}
D.~W. Allan, {\it Proc. IEEE\/} {\bf 54}, 221 (1966).

\bibitem{Howe2008}
D.~Howe, D.~Allan, J.~Barnes, {\it Thirty Fifth Annual Frequency Control
  Symposium\/} (IEEE, 2008), pp. 669--716.

\bibitem{Chekhovich2011}
E.~A. Chekhovich, {\it et~al.\/}, {\it Nat. Nanotechnol.\/} {\bf 7}, 646
  (2012).

\bibitem{Bulutay2012}
C.~Bulutay, {\it Phys. Rev. B\/} {\bf 85}, 115313 (2012).

\bibitem{Bulutay2014}
C.~Bulutay, E.~A. Chekhovich, A.~I. Tartakovskii, {\it Phys. Rev. B\/} {\bf
  90}, 205425 (2014).

\bibitem{Denning2019}
E.~V. Denning, D.~A. Gangloff, M.~Atat{\"{u}}re, J.~M{\o}rk, C.~{Le Gall}, {\it
  Phys. Rev. Lett.\/} {\bf 123}, 140502 (2019).

\bibitem{Bodey2019}
J.~H. Bodey, {\it et~al.\/}, {\it npj Quantum Inf.\/} {\bf 5}, 95 (2019).

\bibitem{Hogele2012}
A.~H{\"{o}}gele, {\it et~al.\/}, {\it Phys. Rev. Lett.\/} {\bf 108}, 197403
  (2012).

\bibitem{Yang2013}
W.~Yang, L.~J. Sham, {\it Phys. Rev. B\/} {\bf 88}, 235304 (2013).

\bibitem{Johansson2013}
J.~Johansson, P.~Nation, F.~Nori, {\it Comput. Phys. Commun.\/} {\bf 184}, 1234
  (2013).

\bibitem{Kuramoto2005}
Y.~Kuramoto, {\it International Symposium on Mathematical Problems in
  Theoretical Physics\/} (Springer-Verlag, 2005), pp. 420--422.

\bibitem{Wust2016}
G.~W{\"{u}}st, {\it et~al.\/}, {\it Nat. Nanotechnol.\/} {\bf 11}, 885 (2016).

\bibitem{Ruderman1954}
M.~A. Ruderman, C.~Kittel, {\it Phys. Rev.\/} {\bf 96}, 99 (1954).

\end{thebibliography}

\noindent
\textbf{Acknowledgements:} We thank Gabriel Éthier-Majcher for helpful discussions. We acknowledge support from  the US Office of Naval Research Global (N62909-19-1-2115), ERC PHOENICS (617985), EPSRC NQIT (EP/M013243/1), EU H2020 FET-Open project QLUSTER (DLV-862035) and the Royal Society (EA/181068). Samples were grown in the EPSRC National Epitaxy Facility. D.A.G. acknowledges a St John's College Fellowship and C.LG. a Dorothy Hodgkin Royal Society Fellowship. 


\end{document}


\title{Quantum sensing of a coherent single spin excitation in a nuclear ensemble: Supplementary information} 

\author{D. M. Jackson \textsuperscript{1*}}
\author{D. A. Gangloff \textsuperscript{1*$\dagger$}}
\author{J. H. Bodey \textsuperscript{1}}
\author{L. Zaporski \textsuperscript{1}}
\author{C. Bachorz \textsuperscript{1}}
\author{E. Clarke \textsuperscript{2}}
\author{M. Hugues \textsuperscript{3}}
\author{C. Le Gall \textsuperscript{1}}
\author{M. Atat\"ure\textsuperscript{1 $\dagger$}}

\noaffiliation

\affiliation{Cavendish Laboratory, University of Cambridge, JJ Thomson Avenue, Cambridge, CB3 0HE, UK}
\affiliation{EPSRC National Epitaxy Facility, University of Sheffield, Sheffield, Broad Lane, S3 7HQ, UK}
\affiliation{Universit\'e C\^ote d'Azur, CNRS, CRHEA, rue Bernard Gregory, 06560 Valbonne, France
\\ \ \\
\textsuperscript{*} These authors contributed equally to this work.
\\
\textsuperscript{$\dagger$} e-mail: dag50@cam.ac.uk; ma424@cam.ac.uk.
\\ \ \\
}

\maketitle


\tableofcontents


\section{Experimental setup}
\label{sec_setup}

\subsection{Quantum-dot device}
\label{subsec_qddevice}

\begin{figure} 
\includegraphics[width = \columnwidth]{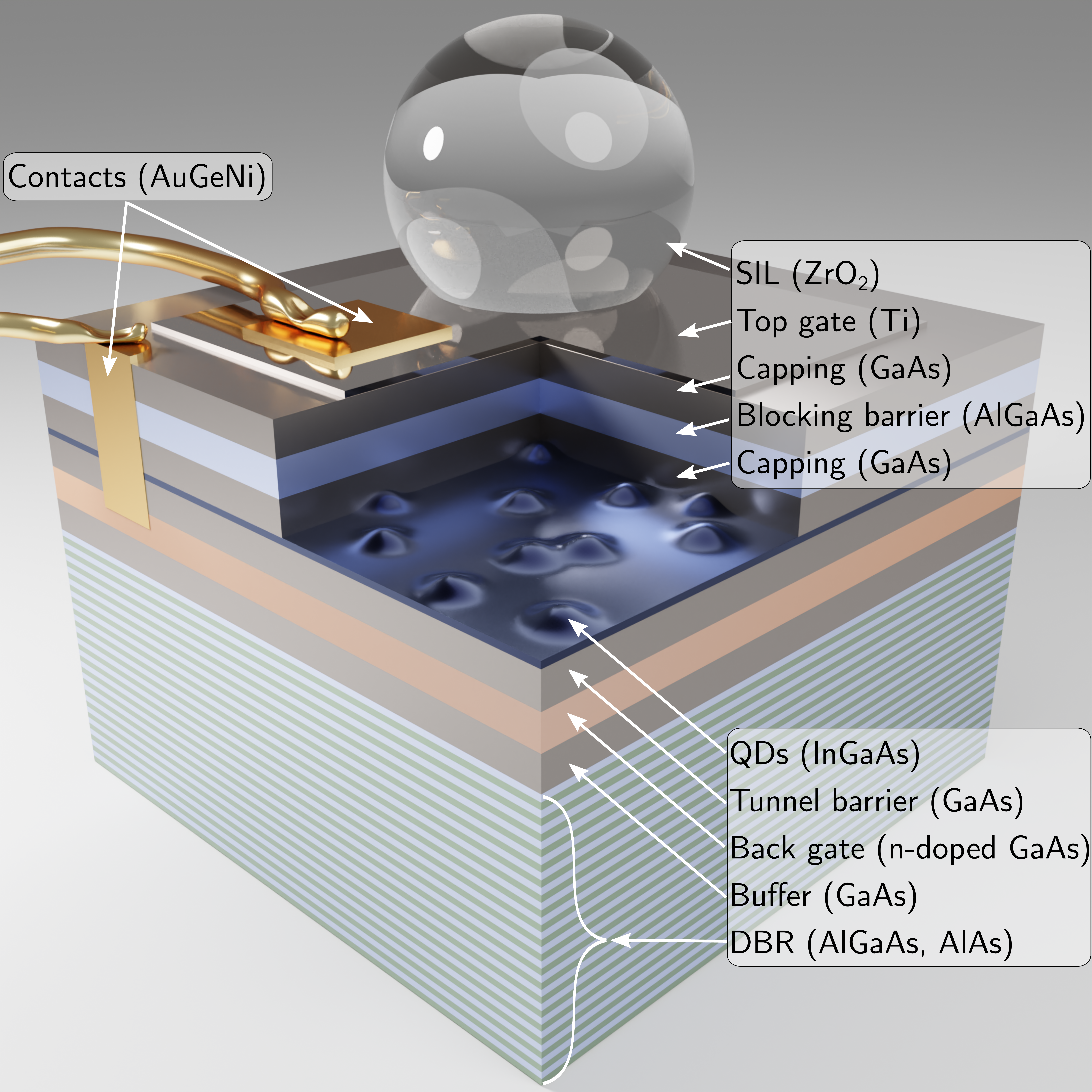}
\caption{\textbf{Sample heterostructure.} A 3D render of the sample heterostructure (not to scale) with a cut-out above the QD layer.}
\label{sample}
\end{figure}

Our quantum-dot (QD) device is from the same wafer as QD devices used in previous works \cite{Stockill2016,Ethier-Majcher2017, Gangloff2019}. Self-assembled InGaAs QDs are grown by Molecular Beam Epitaxy and integrated inside a Schottky diode structure \cite{Urbaszek2013}, above a distributed Bragg reflector (DBR) to maximize photon-outcoupling efficiency (Fig. \ref{sample}). There is a 35-nm tunnel barrier between the n-doped layer and the QDs, and a blocking barrier above the QD layer to prevent charge leakage. The Schottky diode structure is electrically contacted through Ohmic AuGeNi contacts to the n-doped layer, and a semitransparent Ti gate (6 nm) is evaporated onto the surface of the sample. The photon collection is enhanced by the placement of a superhemispherical cubic zirconia solid immersion lens (SIL) on the top Schottky contact of the sample. We estimate a photon-outcoupling efficiency of 10\% for QDs with an emission wavelength around 950 nm.

\subsection{Optical and microwave system}
\label{subsec_schematic}

\begin{figure*}
\centering
\includegraphics[width = 0.8\textwidth]{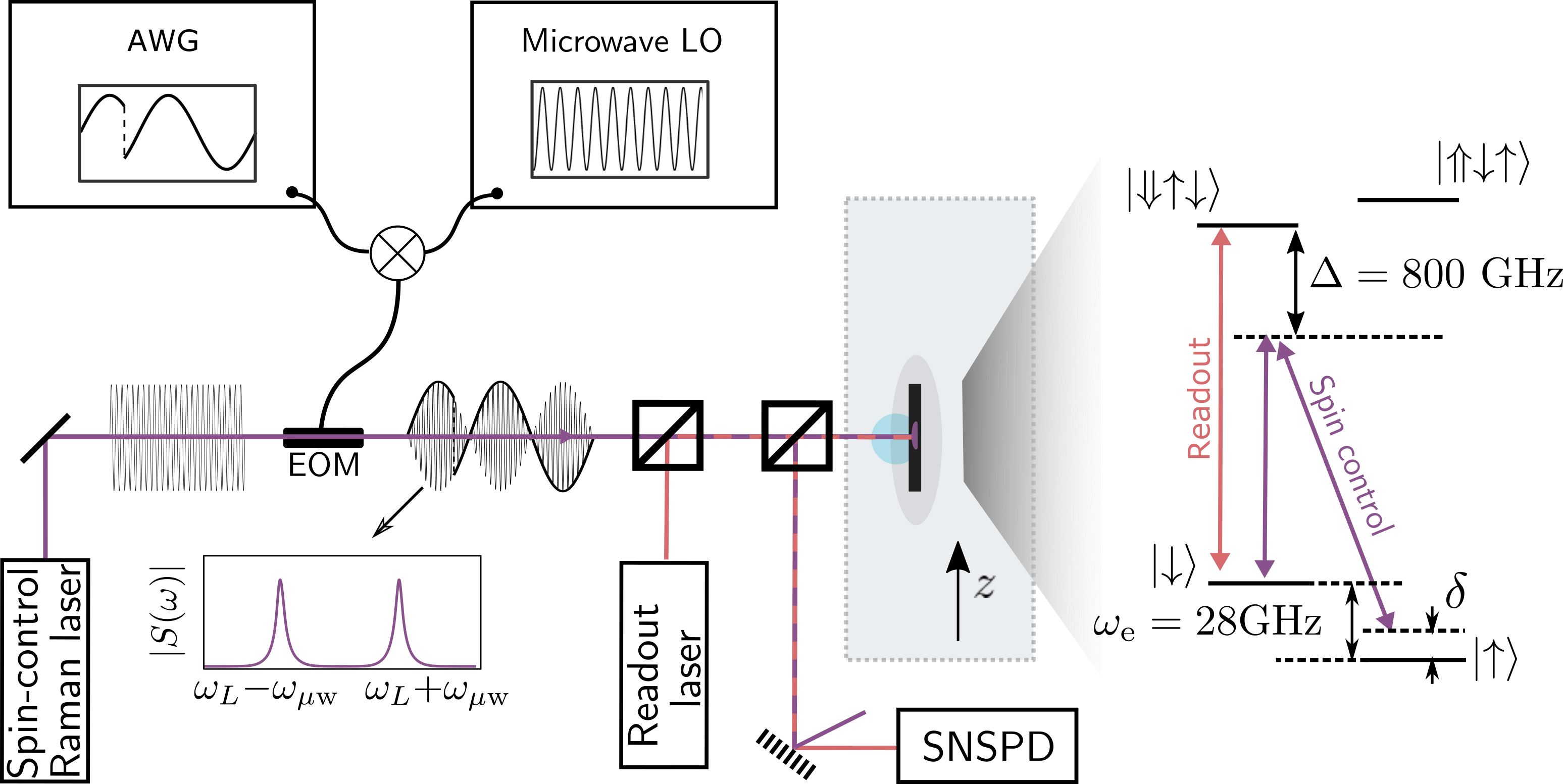}
\caption{\textbf{Experimental schematic.} Amplitude modulation of a single-frequency ($\omega_\text{L}$) laser with an EOM driven by a microwave tone ($\omega_{\mu\text{w}}$) produces two sidebands for spin-control. Encoding a phase step $\Delta\phi_{\mu w}$ in the microwave signal produces a change of relative phase $2\Delta\phi_{\mu w}$ between the two sidebands. These then drive two-photon Raman transitions between the energy levels of a negatively charged QD, as shown on the right. The optical fields have a single-photon detuning from the excited state of $\Delta = 800$\,GHz, and a two-photon detuning from the ESR of $\delta$. Resonant laser pulses optically pump the electron spin at specific moments during the experimental sequence; this serves both to initialize the electron prior to spin control and to read out the population of the spin-$\downarrow$ state. Photons scattered back through the polarization- and frequency-filtered confocal microscope are detected on a Quantum Opus superconducting-nanowire single-photon detector (SNSPD).}
\label{expsetup}
\end{figure*}

A schematic of the experiment is shown in Fig.~\ref{expsetup}. The QD device is placed in a bath cryostat at $4$\,K. A magnetic field is applied transverse to the QD growth axis (Voigt geometry). Two laser systems are combined and sent to the QD: a microwave-modulated Raman laser system and a resonant readout laser (Newport NF laser). A cross-polarization confocal microscope is used for excitation and low-background fluorescence collection; fluorescence also passes through an optical grating filter with a $20$-GHz passband. 

The Raman laser source is a TOPTICA TA Pro. Raman beams are generated by modulating a fibre-based EOSPACE electro-optic amplitude modulator (EOM) with a phase- and amplitude-controlled $14$-GHz microwave tone, constructed as follows. A Rohde \& Schwartz $22$-GHz microwave frequency source generates a fixed amplitude tone at $11$\,GHz, which is frequency mixed (Mini Circuits mixer ZX05-24MH-S+) with the output of a Tektronix arbitrary waveform generator (AWG70001A, $25$\,GSamples/s) operated at $\sim3$\,GHz. The mixer output is high-pass filtered to remove undesired frequency components and retain only a $\sim 14$ GHz tone, and amplified to a peak power of $25$\,dBm. The Raman beams, with a combined optical power of $\sim$$1$\,mW at the device, are passed through a quarter-wave plate and arrive at the quantum dot with near circular polarization at a red single-photon detuning $\Delta \gtrsim 800$\,GHz. The first-order EOM sidebands are two coherent laser fields whose energy difference can be made resonant with the electron spin resonance (ESR), as per this work $28$\,GHz (corresponding to an applied magnetic field $B_z = 4.5$\,T), leading to a two-photon detuning $\delta \approx 0$.

\section{Pulse sequences}
\label{sec_sequences}

\begin{figure*}
\includegraphics[width = \textwidth]{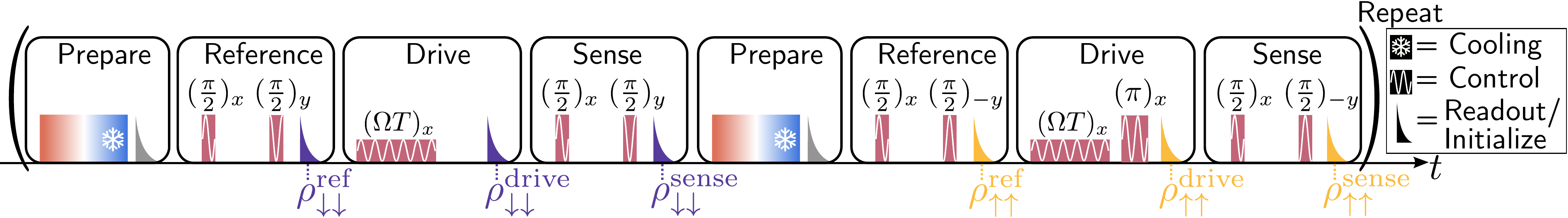}
\caption{\textbf{Experimental pulse sequence.} Full experimental pulse sequence for magnon sensing. Timings are not to scale; preparation (red-blue pulses) typically has a $\sim70$\% duty cycle. The two halves of the sequence differ only in which electron spin population is probed (purple versus yellow readout pulses). Spin rotations on the electron (pink pulses) are labeled with the rotation angle and axis of rotation on the Bloch sphere; the inlaid white curves depict the qubit drive field, and show how steps in its phase can control the axis of rotation.}
\label{pulse_sequence}
\end{figure*}

A single repeating unit of the full experimental pulse sequence is depicted schematically in Fig. \ref{pulse_sequence}. It is split into two very similar sequences, which differ only in the electron spin population that is interrogated. Both are preceded by $11$\,$\mu$s of nuclear spin preparation, unless otherwise stated, equating to a preparation ratio of $\sim 3$. This preparation ensures we always start with a cooled nuclear spin bath. Cooling is performed by driving the electron spin with a Rabi frequency, $\Omega_\text{cool}$, of $13$\,MHz whilst exciting resonantly the optical transition between electron spin-$\downarrow$ and the lowest-energy trion excited state, with a saturation parameter of $0.5$ \cite{Gangloff2019}. This resonant excitation optically pumps the electron, which engineers an effective linewidth for the spin-$\downarrow$ state of $\Gamma_\text{p}=35$\,MHz. 

To interrogate the electronic populations we again optically pump the electron from spin-$\downarrow$ to spin-$\uparrow$ via the excited trion state. We do so for $150$\,ns, which is much longer than the optical pumping time ($\sim 2$\,ns), in order to reach high-fidelity initialization. The resulting brightness, integrated over the first $50$\,ns of this readout pulse, is directly proportional to spin-$\downarrow$ population, $\rho_{\downarrow\downarrow}$. In the second half of the experimental sequence we precede this readout with a spin inversion of the electron (Fig. \ref{pulse_sequence}), yielding a brightness which is instead proportional to spin-$\uparrow$ population, $\rho_{\uparrow\uparrow}$; in this way we can extract both populations and thus electron spin polarization $S=\rho_{\downarrow\downarrow}-\rho_{\uparrow\uparrow}$. Taking the normalized difference of our readouts to obtain spin polarization allows us to cancel any noise that would arise from a drift in outcoupling efficiency.

For an individual control pulse on the qubit we can set its timing, duration, Rabi frequency, phase and detuning. This is thanks to the effective microwave control over the spin qubit provided by our two-photon Raman process. We modify our Raman beams via the power, phase and frequency of the EOM's microwave drive, which in turn control the Rabi frequency, phase and detuning of the qubit drive, respectively. An experimental sequence is thus defined by a microwave signal where all of these parameters, along with pulse timings, are set programmatically with our AWG, at a sampling rate of $25$ GSamples/s.

\section{Calibrations}
\label{sec_calibrations}

\begin{figure*}
\centering
\includegraphics[width = 0.9\textwidth]{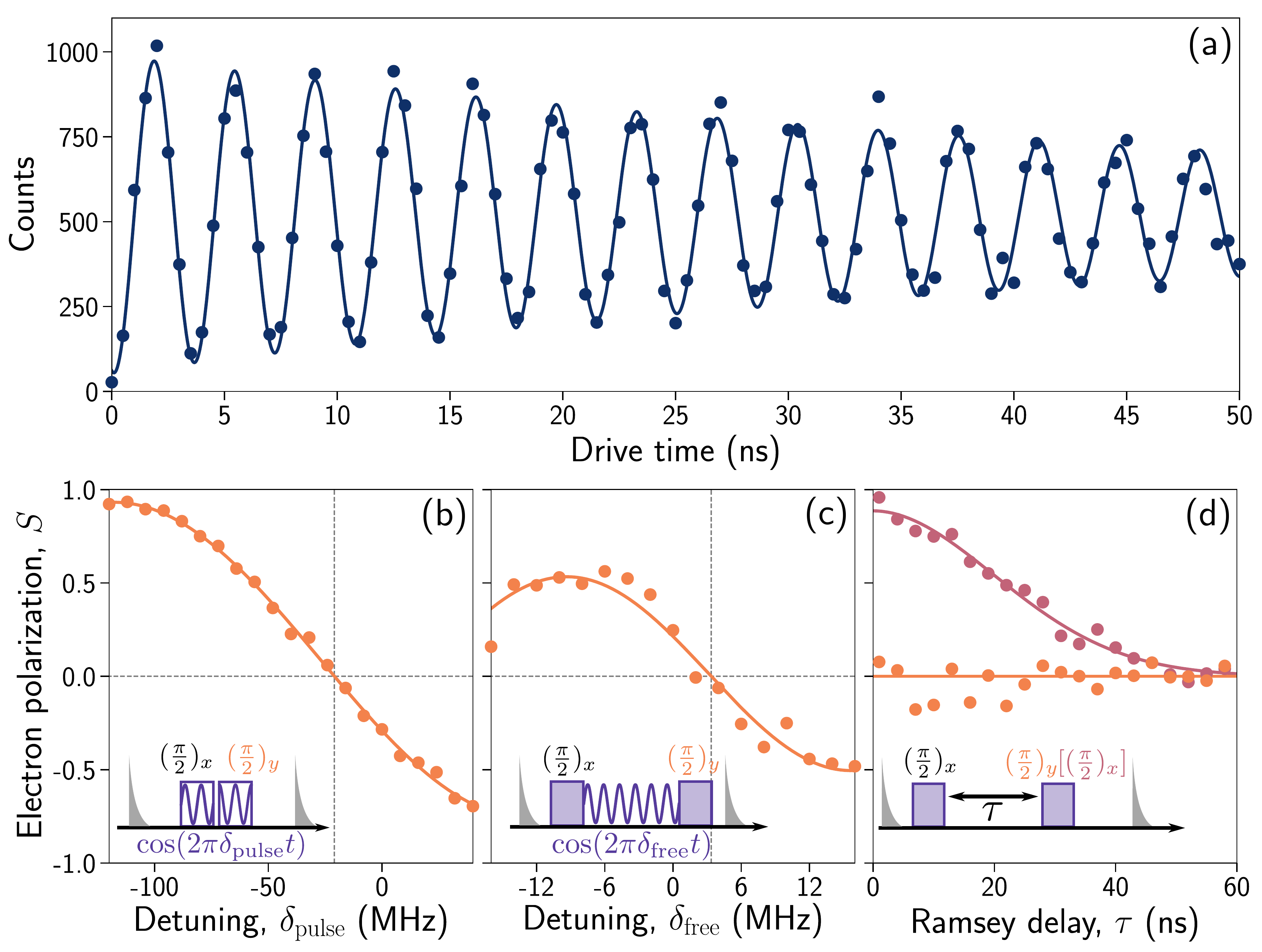}
\caption{\textbf{Ramsey interferometry calibrations.} \textbf{(a)} Rabi oscillations of the electron spin (blue circles) fitted with a mono-exponentially decaying sinusoid (solid curve). These counts are obtained after $4$s of a sequence with a readout duty cycle as in Fig. \ref{pulse_sequence}. From this we extract a Rabi frequency of $\Omega=280$\,MHz and thus calibrate the pulse duration required for $\frac{\pi}{2}$-rotation, i.e. $1/4\Omega=0.89$\,ns. \textbf{(b)} Compensating a systematic pulse error present during each $\frac{\pi}{2}$-rotation in our Ramsey sequence. Electron polarization measured after a \textit{side-of-fringe} Ramsey sequence at zero delay (inset) is plotted versus the correction detuning, $\delta_\text{pulse}$, (orange circles). A sinusoidal fit (solid curve) yields a zero-crossing of $\delta_\text{pulse}\approx-20$\,MHz. \textbf{(c)} Compensating a systematic detuning present during free precession. Electron polarization measured after a \textit{side-of-fringe} Ramsey sequence at finite delay, $\tau\approx 20$\,ns, (inset) is plotted versus the correction detuning, $\delta_\text{free}$, (orange circles). A sinusoidal fit (solid curve) yields a zero-crossing of $\delta_\text{free}\approx-3.5$\,MHz. \textbf{(d)} Ramsey interferometry signals after careful calibration for both \textit{top-of-fringe} (pink circles) and \textit{side-of-fringe} (orange circles) configurations. The former is fitted to a Gaussian free-induction decay (pink curve), from which we obtain the characteristic decay time $T_2^*\approx30$\,ns. In the latter, electron spin polarization is pinned to $0$ (orange curve), i.e. the side of a Ramsey fringe.}
\label{calibrations}
\end{figure*}

Before any experiment we calibrate the Ramsey control sequence used for magnon sensing. Specifically we aim to configure the Ramsey interferometer in the ideal \textit{side-of-fringe} configuration, which is maximally sensitive to small magnon-induced ESR frequency shifts. Firstly, we calibrate the qubit-drive pulse duration required in order to perform a $\frac{\pi}{2}$-rotation of the electron spin, which is simply a quarter of a Rabi period. To this end we drive Rabi oscillations of the qubit for many periods (Fig. \ref{calibrations}a) to extract the Rabi frequency, $\Omega=280$\,MHz, and thus the gate time, $1/4\Omega=0.89$\,ns. Secondly, we calibrate the drive frequency used for the $\frac{\pi}{2}$-rotations. In the ideal case, we simply set this to the bare ESR frequency, meaning all rotations occur in a frame rotating with the qubit. In reality, there are two systematic pulse errors that we calibrate and correct using appropriate drive detunings, as detailed below. In this way we ensure that the rotating frame of the drive matches the qubit at all times.

\subsection{Pulse compensation}
\label{subsec_pulse_compensation}

The high optical power used during a $\frac{\pi}{2}$-pulse leads to a differential optical Stark shift on the two spin eigenstates. If left unchecked, this effective detuning of the qubit drive leads to rotations which are not about the desired axes. We compensate this effect by changing the drive frequency used during the pulse by $\delta_\text{pulse}$ to match the shifted qubit frequency. To find the correct $\delta_\text{pulse}$ we use a \textit{side-of-fringe} Ramsey sequence with zero Ramsey delay ($\tau=0$), i.e. a $(\frac{\pi}{2})_x$ rotation followed immediately by a $(\frac{\pi}{2})_y$ rotation (Fig. \ref{calibrations}b inset). At the zero-crossing of electron polarization, $S$, versus $\delta_\text{pulse}$ (orange circles), we have applied the correct detuning, $\delta_\text{pulse}\approx-20$\,MHz, to configure the Ramsey interferometer on the \textit{side-of-fringe}.

\subsection{Free-precession compensation}
\label{subsec_rotating_frame}
{}
The second systematic pulse error is a detuning originating from the nuclear-preparation step of the experimental sequence (Fig. \ref{pulse_sequence}). We engineer feedback on the nuclear polarization in order to lock the ESR frequency to that of our qubit drive. With the drive frequency set equal to the bare ESR, this defines a lockpoint around zero nuclear polarization, $\langle I_z \rangle\approx0$ \cite{Gangloff2019}. Crucially, part of this feedback requires resonant pumping of the optical transition, as in Fig. \ref{expsetup}, which optically Stark shifts the lockpoint away from the drive frequency by a few megahertz. Such a shift between the locked ESR frequency and the drive is precisely what we detect during the free precession of a Ramsey sequence. Indeed it would be added to any magnon-induced ESR shift we hope to measure, obfuscating single-magnon detection. To this end, we cancel the optically induced ESR shift with an opposite detuning on the qubit drive, $\delta_\text{free}$, during free precession. Since the drive is off during this time, a detuning is instead applied as a phase, $2\pi\delta_\text{free}\tau$, to the second Ramsey pulse. To find the correct $\delta_\text{free}$ we use a \textit{side-of-fringe} Ramsey sequence with a finite delay (Fig. \ref{calibrations}c inset), again looking for the zero crossing of electron polarization versus $\delta_\text{free}$. At this point the rotating frame defined by the drive frequency is matched precisely to the qubit. This means that as we scan $\tau$ in Fig. \ref{calibrations}d (orange circles) there is no phase accumulation of the qubit, and electron polarization is pinned to the \textit{side-of-fringe}.

\section{Ramsey measurements}
\label{sec_ramsey}

\subsection{Optimal Ramsey sequence}
\label{subsec_optimal}

Having calibrated our Ramsey interferometer to the \textit{side-of-fringe} configuration, we expect the signal in electron polarization to obey the following equation:
\begin{equation}
S^\text{sense}=\rho_{\downarrow\downarrow}^\text{sense}-\rho_{\uparrow\uparrow}^\text{sense} = e^{-\frac{\tau}{T_2^*}^2}\sin(2\pi\tau\Delta\omega_\text{sense}).
\label{sof_ramsey}
\end{equation}
Here we map a small phase angle, $2\pi\tau\Delta\omega_\text{sense}$, to a change in electron polarization linearly, compared to a \textit{top-of-fringe} configuration, which goes as $\cos(2\pi\tau\Delta\omega_\text{sense})$, where there is a less sensitive quadratic mapping. Both situations are depicted in Fig. \ref{calibrations}d for $\Delta\omega_\text{sense}=0$. In the latter configuration we can extract the characteristic decay time $T_2^*\approx30$\,ns, which sets the maximum Ramsey delay we can use for sensing before the qubit loses its coherence. During this time the phase accumulation we expect due to a $\sim200$\,kHz ESR shift is only $\sim0.01\pi$. For small angles such as this, a linear mapping to electron polarization is clearly advantageous. To a good approximation the expected signal is then $S^\text{sense}\approx e^{-(\tau/T_2^*)^2}2\pi\tau\Delta\omega_\text{sense}$, which has a maximum at $\tau=T_2^*/\sqrt{2}$. As such we conclude that a Ramsey interferometer configured on the \textit{side-of-fringe} and fixed at this delay is the most sensitive sequence for detecting sub-$200$\,kHz ESR shifts \cite{Degen2017}.

\subsection{Sensor performance: Allan deviation}
\label{subsec_allan}

\begin{figure}
\includegraphics[width = \columnwidth]{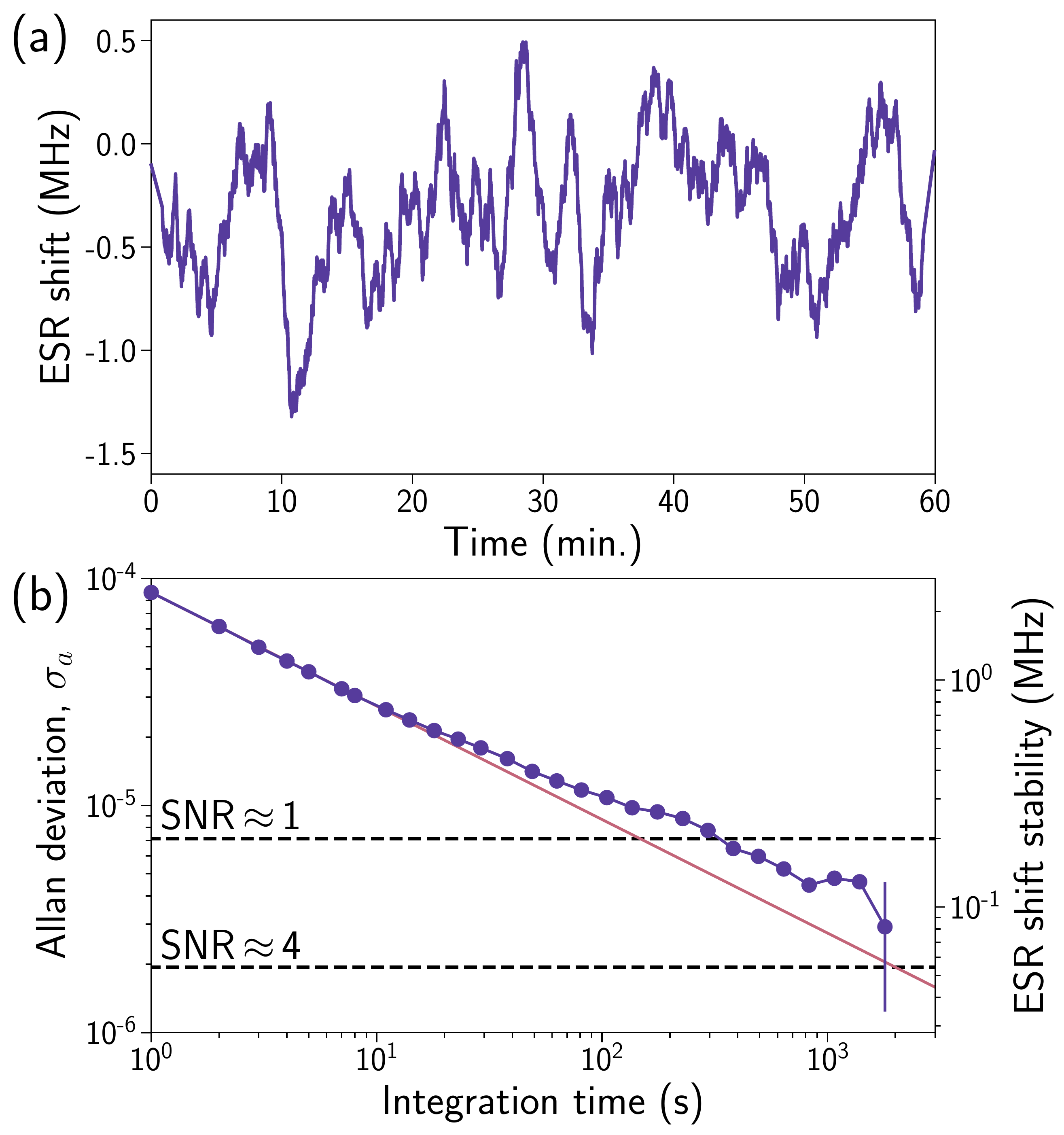}
\caption{\textbf{Allan deviation of ESR frequency.} \textbf{(a)} Time trace of repeated measurements of the ESR shift, $\Delta\omega_\text{ref}$, (purple curve). \textbf{(b)} Purple curve is the overlapping Allan deviation, $\sigma_a$, calculated from the above time-series, as a function of integration time; i.e. the time over which we average the measured ESR shift. We multiply the dimensionless $\sigma_\text{a}$ by the qubit frequency, $\omega_\text{e}$, to obtain the absolute ESR-frequency stability on the right axis. The single-magnon ESR shift is $\sim200$\,kHz, which defines the signal-to-noise ratio (SNR) thresholds depicted. The extrapolated short-time behavior (pink curve) scales inversely with the square root of the integration time.}
\label{fig_allan_deviation}
\end{figure}

We characterize sensor performance by using our precision Ramsey sequence to make repeated measurements of the \textit{reference} ESR shift, $\Delta\omega_\text{ref}$, at a sampling rate of $1$ Hz. The time-trace of these measurements over the course of one hour is presented in Fig. \ref{fig_allan_deviation}a, where we have smoothed the data with a $100$ second moving average. Since there is no magnon injection we expect to measure no shift from the bare electron spin splitting, and so an average $\Delta\omega_\text{ref}=0$. There is, however, a small $\sim500$\,kHz error in the $\delta_\text{free}$ calibration, which has shifted the mean; this has no consequence for the following discussion as we only seek to quantify the variations in $\Delta\omega_\text{ref}$. From the time trace in Fig. \ref{fig_allan_deviation}a we can see qualitatively that there are slow drifts in the ESR shift on top of higher frequency noise. To quantify this we calculate the Allan deviation from our series of $m=3600$ samples $\{\Delta\omega^\text{ref}_i\}$ \cite{Allan1966}, which we first divide by the qubit frequency, $\omega_\text{e}$, to obtain dimensionless, fractional frequencies $\{f^\text{ref}_i\}$. For an integration time of $n$ seconds, the overlapping Allan variance \cite{Howe2008} is then given by
\begin{equation}
\sigma_a^2(n)=\frac{1}{2n^2(m-2n+1)}\sum_{j=0}^{m-2n}\left(\sum_{i=j}^{j+n-1}{f^\text{ref}_{i+n}-f^\text{ref}_{i}}\right)^2.
\label{equ_allan_var}
\end{equation}
The square-root of this quantity\textemdash plotted in \ref{fig_allan_deviation}b as a function of the averaging time (purple circles)\textemdash is a metric of the fluctuations in the qubit frequency. We are characterising fluctuations in the ESR frequency with precisely the method we use to detect the single magnon ESR shift. As such this metric captures all the noise sources we can expect to limit the precision of an ESR shift measurement. Practical limitations mean that we can only operate with an averaging time of up to $2000$ seconds. In this regime fluctuations in ESR shift are twice what would be expected from the quantum-limited noise of our photonic readout, which is given by the extrapolation of the short averaging-time behaviour. These low-frequency additional noise sources could be both physical and technical. For example there may be slow drifts in nuclear polarization on the $100$\,kHz scale that are not eliminated by nuclear spin preparation. Additionally noise could stem from slow drifts in optical power, altering the systematic detunings, $\delta_\text{pulse}$ and $\delta_\text{free}$, that were calibrated for in section \ref{sec_calibrations}. Regardless of the source, we outline in the main text how these slow drifts can be eliminated with a differential measurement. 

\section{Modelling magnon dynamics as simple harmonic motion}
\label{sho_derivation}

\begin{figure}
\includegraphics[width=\columnwidth]{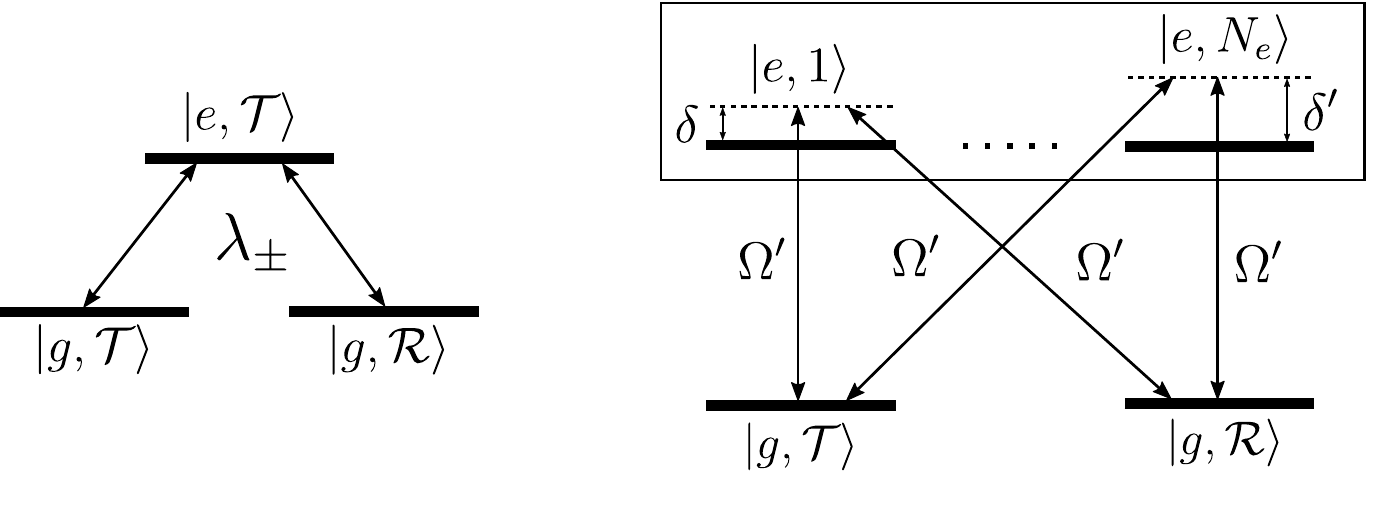}
\caption{\textbf{Magnon dynamics as a two-level system.} Energy level diagram representing the two-level system $\{\ket{g,\mathcal{T}},\ket{e,\mathcal{T}}\}$ used to model simplified magnon dynamics. The excited state is $N$-fold degenerate, each with a randomly distributed detuning $\delta_e$. $\Omega'$ is the single-nucleus Rabi frequency, which couples each degenerate excited state to the ground state. $\ket{g,\mathcal{R}}$ is a single state which models the leakage of population out of the target manifold $\mathcal{T}$.}
\label{fig_inhomog_energy_diagram}
\end{figure}

We wish to model the injection of a magnon into $N$ nuclear spins as the drive of a two level system with a non-degenerate ground state $\ket{g,\mathcal{T}}$ and an $N$-fold degenerate excited state $\ket{e,\mathcal{T}}$ (Fig. \ref{fig_inhomog_energy_diagram}): an exact model for a fully polarized nuclear ensemble. Presently, we neglect an additional state $\ket{g,\mathcal{R}}$ represented in Fig. \ref{fig_inhomog_energy_diagram}, which later will come to represent the leakage of population from the target manifold $\mathcal{T}$ to an undesired residual manifold $\mathcal{R}$. The single-nucleus Rabi frequency is given by $\Omega'=\alpha\Omega$ where $\Omega$ is the qubit-drive Rabi frequency. Working in the rotating frame we then find that the probability amplitudes $a_g$ and $\{a_e\}_{1..N}$ for being in the ground and excited states, respectively, are captured by the following equations:
\begin{equation}
\begin{split}
i\dot{a}_e &= \Omega^\prime a_g e^{i\delta_e t}\\
i\dot{a}_g &= \Omega^\prime \sum_e a_e e^{-i\delta_e t}
\end{split}
\label{equ_neff_system}
\end{equation}
where we have made the rotating wave approximation. Strain inhomogeneities in a QD couple with the quadrupolar Hamiltonian to produce an inhomogeneous distribution of nuclear resonances \cite{Chekhovich2011,Bulutay2012,Bulutay2014}, captured here by a randomly distributed $\delta_e$ (Fig. \ref{fig_inhomog_energy_diagram}).

Equations \ref{equ_neff_system} can be decoupled, and expressed through an integro-differential equation after taking $a_g(0)=1$:
\begin{equation}
\dot{a}_g= -\Omega'^2 \sum_e \int_0^t\, dt^\prime  a_g(t^\prime)e^{i\delta_e (t^\prime-t)}
\label{equ_integral_solution}
\end{equation}
In the asymptotic $N$-limit, $\delta_e$ is described by a spectral distribution $S(\delta)$, where $\int d\delta S(\delta) = N$, and equation \ref{equ_integral_solution} takes the form:
\begin{equation}
\dot{a}_g \xrightarrow{N\to \infty}- \Omega^{\prime 2} \int_0^t \, dt^\prime a_g(t^\prime)
\int_{-\infty}^\infty \,d\delta S(\delta)e^{i\delta(t^\prime-t)}
\label{equ_integro_diff}
\end{equation}
Assuming a Lorentzian distribution for $S(\delta)$, with a full width at half maximum of $\Delta$, simplifies equation \ref{equ_integro_diff} yielding
\begin{equation}
\dot{a}_g = -N \Omega^{\prime 2} \int_0^t \, dt^\prime a_g(t^\prime) \exp{\Delta(t^\prime-t)}
\label{equ_neff_simplified}
\end{equation}
This is a standard example of a Delay Differential Equation, which can be solved exactly. Differentiating equation \ref{equ_neff_simplified} using the Leibnitz rule, we arrive at a linear second order differential equation:
\begin{equation}\label{neff_2ndordersys}
\ddot{a}_g=-N \Omega^{\prime 2} a_g -\Delta \dot{a}_g 
\end{equation}
which we immediately recognize as that which governs simple harmonic motion. This can be solved easily using the ansatz $a_g\propto e^{\lambda t}$, yielding:
\begin{equation}
\lambda_\pm = -\frac{\Delta}{2} \pm \sqrt{\frac{\Delta^2}{4}-N\Omega^{\prime 2}}
\end{equation}
In this way we have simplified the dynamics of magnon injection to damped simple harmonic motion with a damping rate $\Gamma=\Delta/2$ and a collectively enhanced drive strength $\Omega_\text{exc}=\sqrt{N}\Omega'$.

At this point point we note that the above solution predicts unphysically that the probability of being in the ground state at infinite time is zero, as opposed to $1/2$. To rectify this we now return to consider the additional ground state $\ket{g,\mathcal{R}}$ in Fig. \ref{fig_inhomog_energy_diagram}, which captures approximately the leakage of population into the manifold $\mathcal{R}$ consisting of all other consistent ground states in the system. This leakage will be at its maximum in our case since we operate around zero nuclear polarization \cite{Denning2019}. With this extension to the above calculation we again recover simple harmonic motion, but crucially we find that the probability to be in this now two-fold degenerate ground state manifold decays to $1/2$ as desired.

\section{Extended data}
\label{sec_data}

\subsection{ESR-shift spectra}
\label{subsec_dnp}

\begin{figure}
\centering
\includegraphics[width = \columnwidth]{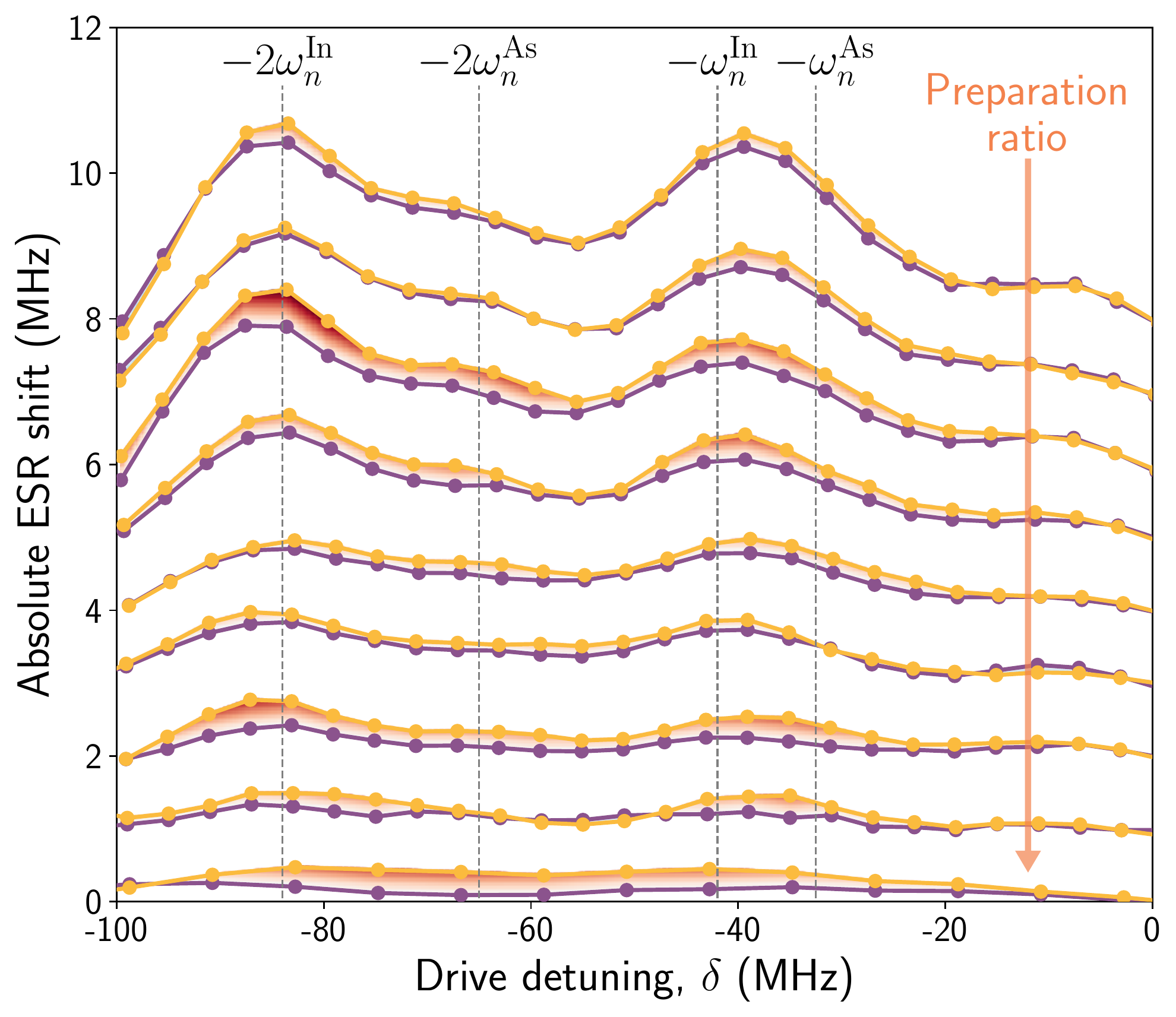}
\caption{\textbf{ESR shift spectra.} The absolute value of the ESR-shift spectra (negative drive detunings only) measured during \textit{reference} and \textit{sense} sequences (purple and yellow circles, respectively) for varying preparation ratio. Preparation-ratio ranges from $1.4$ (top) to $27.5$ (bottom). Solid curves are linear interpolations between data points. Spectra have been offset by $1$\,MHz for clarity.}
\label{fig_dnp_spectra}
\end{figure}

In Figs. 3c and 3d of the main text we present the ESR shifts measured after driving the $I_z-2$ and $I_z-1$ magnon modes, respectively. Specifically, we plot these shifts, whose magnitude we attribute to dynamic nuclear polarization (DNP) that accumulates from one cycle of the pulse sequence to the next, versus preparation ratio, $r=T_\text{prep}/T_\text{exp}$. These shifts have been extracted from the fitted amplitudes of the corresponding spectral features of Fig. \ref{fig_dnp_spectra}, where we offset the spectra for clarity. Each spectrum is obtained after $1.8$\,$\mu$s of qubit drive with a Rabi frequency of $8$\,MHz. For the $I_z-2$ mode we can extract the amplitude of the spectrally resolved $-2\omega_\text{n}^\text{In}$ feature, whereas for the $I_z-1$ mode we can only fit the unresolved $-\omega_\text{n}^\text{In}$/$-\omega_\text{n}^\text{As}$ feature. Nevertheless, in Fig. \ref{fig_dnp_spectra} we see explicitly the quenching of DNP as we increase the fraction of time spent cooling the nuclear ensemble. As such, for the largest preparation ratio, the spectrum of ESR shift measured prior to exciting a magnon, $\Delta\omega_\text{ref}$, approaches zero, and that measured after, $\Delta\omega_\text{sense}$, approaches the single-magnon spectrum. In Figs. 3c and 3d of the main text we capture the $r$-dependence of the measured ESR shifts with the following phenomenological functional form:

\begin{align}
\begin{split}
	&\Delta\omega_\text{ref}(r;a,b)=\frac{a}{1+r/b}\\
	&\Delta\omega_\text{sense}(r;a,b,c)=\Delta\omega_\text{ref}(r;a,b)+c
\end{split}
\label{equ_dnp_func}
\end{align}
\noindent
We optimize $a,b$ and $c$ for $\Delta\omega_\text{ref}$ and $\Delta\omega_\text{sense}$ simultaneously with a least-squares fitting algorithm. The resulting parameters are $a=15(8)$\,MHz, $b=0.3(2)$, $c=210(50)$\,kHz for the $I_z-1$ mode and $a=9(3)$\,MHz, $b=0.6(3)$, $c=200(70)$\,kHz for the $I_z-2$ mode.

\subsection{Magnon oscillations}
\label{subsec_osc}

\begin{figure}
\centering
\includegraphics[width = \columnwidth]{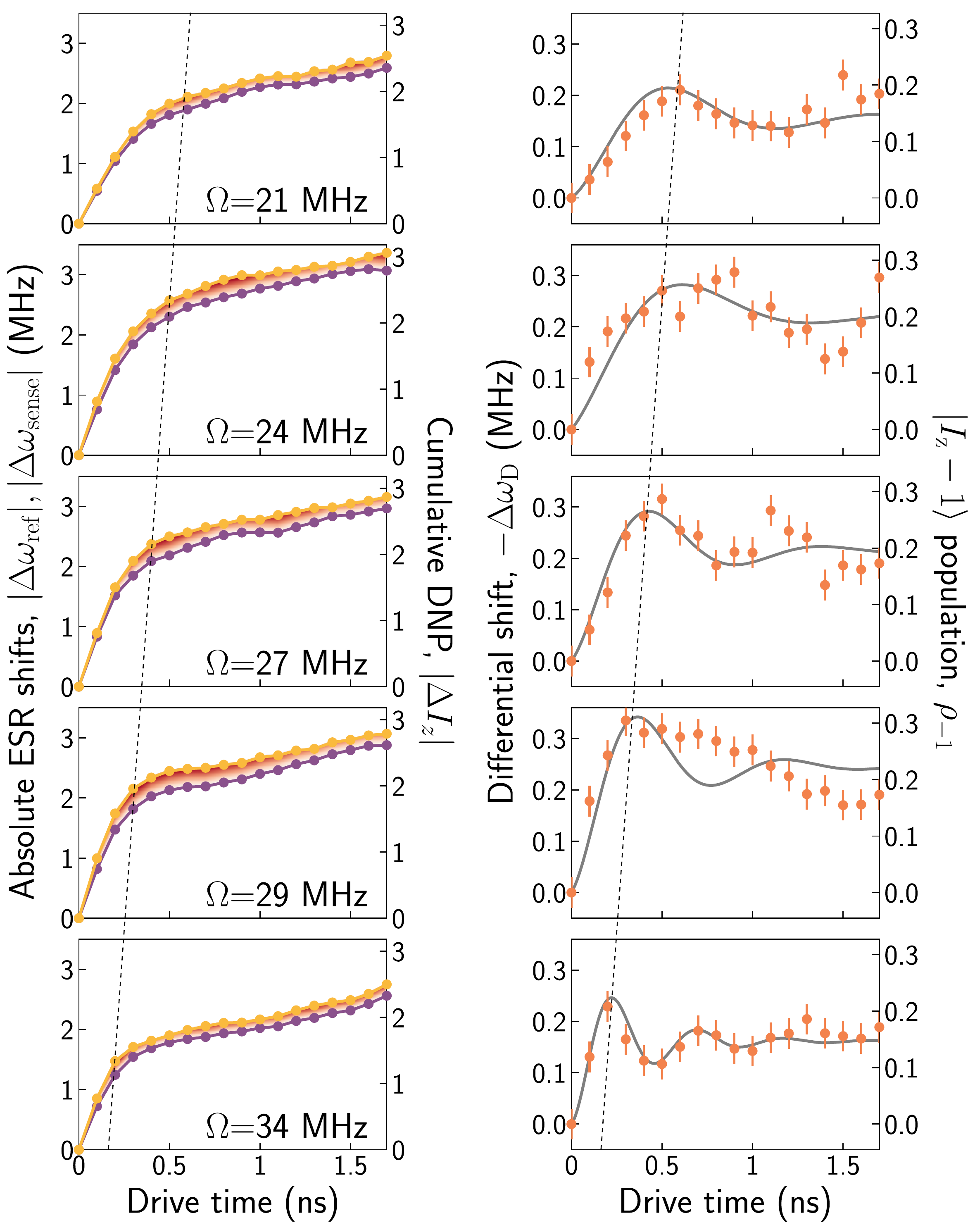}
\caption{\textbf{Rabi oscillations of $\ket{I_z}\leftrightarrow \ket{I_z-1}$ transition.} \textbf{(left)} Absolute value of the ESR shifts, $\abs{\Delta\omega_\text{ref}}$ and $\abs{\Delta\omega_\text{sense}}$, measured during time-resolved driving of the $\ket{I_z}\leftrightarrow \ket{I_z-1}$ transition (purple and yellow circles, respectively). To convert these shifts to the corresponding accumulated DNP (right axis) we dividing by $2A$. \textbf{(right)} The difference of these two shifts, $\Delta\omega_\text{D}$, is the single magnon ESR shift (orange circles). We plot its negative here for clarity, and also convert to population in the $\ket{I_z-1}$ state, $\rho_{-1}=-\Delta\omega_\text{D}/2A$, (right axis). We fit the time evolution to that of a damped harmonic oscillator (solid curves). The dotted lines are a guide to the eye, depicting the approximate $\pi$-time of the electron-magnon exchange for each drive Rabi frequency, $\Omega$.}
\label{fig_dnp_oscillations_fsb}
\end{figure}

\begin{figure}
\includegraphics[width = \columnwidth]{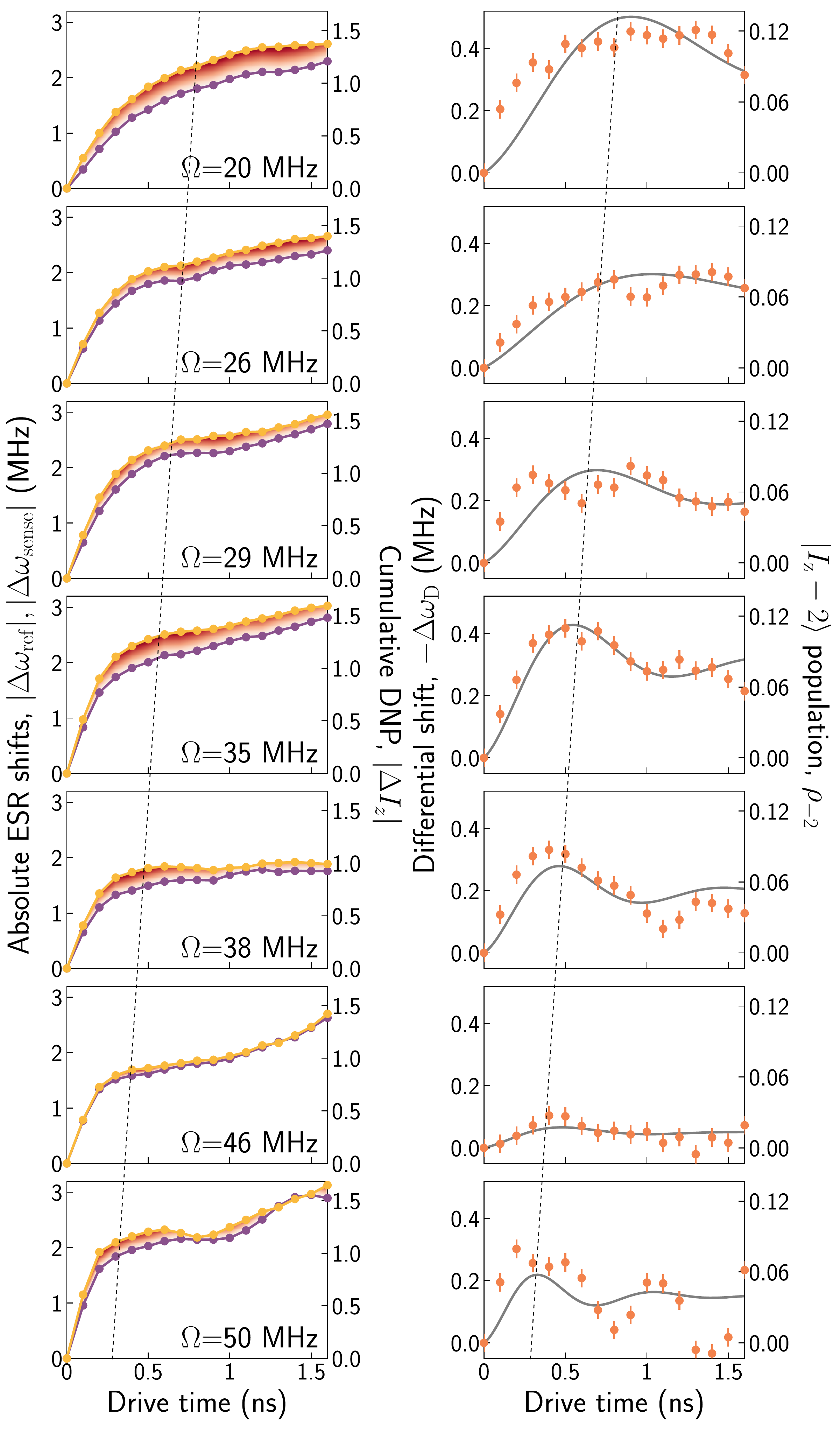}
\caption{\textbf{Rabi oscillations of $\ket{I_z}\leftrightarrow \ket{I_z-2}$ transition.} \textbf{(left)} Absolute ESR shifts, $\abs{\Delta\omega_\text{ref}}$ and $\abs{\Delta\omega_\text{sense}}$, as in Fig. \ref{fig_dnp_oscillations_fsb}, but now for time resolved driving of the $\ket{I_z}\leftrightarrow \ket{I_z-2}$ transition. \textbf{(right)} The negative differential shift, $-\Delta\omega_\text{D}$, as in Fig. \ref{fig_dnp_oscillations_fsb}, except we instead must calculate population in the $\ket{I_z-2}$ state, $\rho_{-2}=-\Delta\omega_\text{D}/4A$.}
\label{fig_dnp_oscillations_ssb}
\end{figure}

In Figs. \ref{fig_dnp_oscillations_fsb} and \ref{fig_dnp_oscillations_ssb} we present supplementary data to Fig. 4 of the main text, which depict the time-resolved ESR shifts under resonant driving of the $I_z-1$ and $I_z-2$ modes, respectively. For multiple drive Rabi frequencies we plot both the cumulative and differential shifts, the former being enhanced by the accumulation of DNP because we operate with a preparation ratio of only $3$. For both magnon modes we divide these cumulative ESR shifts by $2A$ to convert to the corresponding DNP, $\Delta I_z$. Similarly, we also convert the differential shift to magnon population in the $\ket{I_z-1}$ and $\ket{I_z-2}$ states by dividing by $2A$ and $4A$ respectively. To obtain the drive Rabi-frequency dependence of the coherent dynamics we fit the oscillatory behaviour of $\Delta\omega_\text{D}$ to that of a damped harmonic oscillator, as per section \ref{sho_derivation}:
\begin{equation}
\Delta\omega_\text{D} = \omega_0\left[1-\frac{e^{-\Gamma T}}{2}\left(e^{iT\sqrt{\Omega_\text{exc}^2-\Gamma^2}}+e^{-iT\sqrt{\Omega_\text{exc}^2-\Gamma^2}}\right)\right].
\label{equ_fit_func}
\end{equation}
We leave the fitting algorithm free to explore both the over- and under-damped regimes, but as can be seen from Figs. \ref{fig_dnp_oscillations_fsb} and \ref{fig_dnp_oscillations_ssb} it returns under-damped motion as the best fit in all cases. The fitted exchange frequency, $\Omega_\text{exc}/2\pi$, is what is plotted in Fig. 4d of the main text.

\section{Modelling the electron-nuclear interface}
\label{sec_model}

\subsection{System Hamiltonian}
\label{subsec_hamiltonian}{}

\begin{figure*}
\includegraphics[width = 0.85\textwidth]{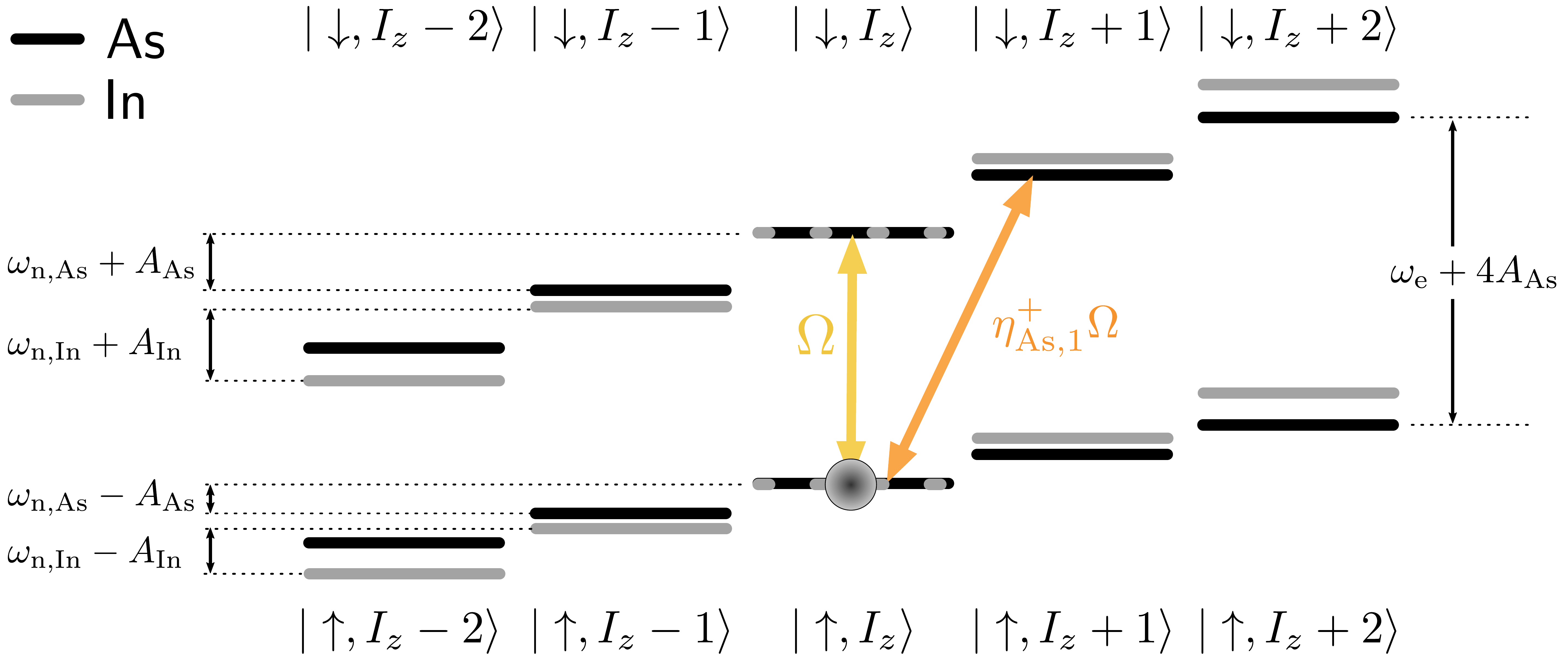}
\caption{\textbf{Magnon level scheme.} Lab-frame eigenstates of the undriven electron spin dressed with collective nuclear states, $I_z$, of arsenic (black) and indium (grey). Yellow arrow represents qubit drive with Rabi frequency $\Omega$, which in turn activates the magnon injection (orange arrow) thanks to the quadrupolar-enabled noncollinear hyperfine interaction. The ladder of states is anharmonic since the dressed electronic excited (ground) state shifts by $\omega_{\text{n},j} I_{z,j}+A_j I_{z,j}$ ($\omega_{\text{n},j}I_{z,j}-A_j I_{z,j}$), for nuclear species $j$.}
\label{fig_level_scheme}
\end{figure*}

We consider the following system Hamiltonian \cite{Gangloff2019,Denning2019,Bodey2019} representing an electron spin qubit and the collective states of two nuclear species, arsenic (As) and indium (In):

\begin{align}
\begin{split}
  H_\text{tot} = &\delta S_z + \Omega S_x + \omega_\text{n,As} I_{z,\text{As}} + \omega_\text{n,In} I_{z,\text{In}} \\
  &- 2 S_z (A_\text{As} I_{z,\text{As}} + A_\text{In} I_{z,\text{In}} )\\
  &+ S_z \sum_j \sum_k A_{\text{nc},j,k} \left(M_{j,k}^+ + M_{j,k}^-\right)
\end{split}
\label{equ_szham}
\end{align}

The terms $\delta S_z + \Omega S_x$ are control of the electronic spin $S_i = \sigma_i/2$ ($\sigma_i$ are the Pauli operators), which is split by $\omega_\text{e}$ by an external magnetic field $B_z$, and addressed in the spin's rotating frame via a resonant drive with Rabi frequency $\Omega$ and detuning $\delta$ (we have made the rotating-wave approximation); the drive comes from an optically stimulated Raman process \cite{Gangloff2019,Bodey2019}. The collective states of the nuclear spins are captured by the projection of their total angular momentum along the external field, i.e. their polarizations $I_{z,\text{As}}$ and $I_{z,\text{In}}$. The Zeeman effect on the nuclei is represented by $\omega_\text{n,As} I_{z,\text{As}} + \omega_\text{n,In} I_{z,\text{In}}$, where $\omega_\text{n,As}$ and $\omega_\text{n,In}$ are the nuclear Zeeman frequencies for As and In, respectively. The contact hyperfine interaction between the electron and the nuclei takes the form $-2 S_z (A_\text{As} I_{z,\text{As}} + A_\text{In} I_{z,\text{In}} )$, where $A_\text{As}$ and $A_\text{In}$ are the hyperfine constants per nuclear spin -- here we have ignored the transverse part of the hyperfine interaction which at high magnetic field is suppressed by $1/\omega_\text{e}$, and we have considered a homogeneous interaction between the electron and the nuclear spins. The energy level diagram (in the lab frame) is shown in Fig.~\ref{fig_level_scheme}.

The non-collinear Hamiltonian $A_{\text{nc},j,k} S_z M^\pm_{j,k}$ arises from dressing the hyperfine term $-2S_zA_jI_{z,j}$ with strain-induced quadrupolar nuclear effects\cite{Hogele2012, Urbaszek2013, Gangloff2019, Denning2019, Bodey2019}. The operators $M^\pm_{j,k}$ allow a change in the collective nuclear projection $I_{z,j}$ of species $j$ -- nuclear magnon (spin-wave) modes of two polarities ($\pm$) and two types (single $k=1$ or double $k=2$) -- when coupled with an electron spin flip in the driven (dressed-state) basis; in other words a Hartmann-Hahn resonance $\Omega^2 + \delta^2 = \omega_{\text{n},j}^2$. The strength of this interaction is captured by the energy scale $A_{\text{nc},j,k} = \sqrt{g(I_j)N_j} \left(A_jB_{\text{Q},j}/\omega_{\text{n},j}\right) f(\theta_j)$, where we have used a $\sqrt{N_j}$ collective enhancement according to the number of nuclei of each species $N_j$ \cite{Yang2013}. $B_{\text{Q},j}$ and $\pi/2-\theta_j$ are the quadrupolar interaction energy and angle relative to the external magnetic field.  The scaling factor $g(I_j)$ is a function of the total spin of each species, $g(I_\text{As}=3/2)=6$ and $g(I_\text{In}=9/2)=1584/5$, and is derived from the quadrupolar nature of the magnon operators \cite{Bodey2019}. $f$ is a trigonometric function of the quadrupolar angle $\theta_j$. The Hamiltonian in Eq. \ref{equ_szham} is the one used for numerical simulations. 

\subsubsection{Approximate Hamiltonian}

In the main text, we have simplified the Hamiltonian by further dressing the driven electron with the non-collinear Hamiltonian \cite{Hogele2012,Gangloff2019}, which is valid in proximity of the Hartmann-Hahn resonance, $\delta^2 + \Omega^2 \approx \omega_{\text{n},j}^2$. This yields the electron-activated exchange term $-\sum_{j,k} \eta_{j,k} \Omega S_y M^{\pm}_{j,k}$, where the driven electron is more explicit (i.e. $\Omega S_y$). The simplified Hamiltonian, per nuclear species $j$, then becomes:

\begin{align*}
\begin{split}
  H = &\delta S_z + \Omega S_x + \omega_{\text{n},j} I_{z,j} \\
  &- 2 S_z A_j I_{z,j} \\
  &- \sum_{k=1}^{2} \eta_{j,k} \Omega S_y M^{\pm}_{j,k}
\end{split}
\end{align*}

For clarity, the Hamiltonian of equation $1$ in the main text considers only a single nuclear species, thus suppressing index $j$, and by assuming proximity to the resonance $\delta^2 + \Omega^2 \approx \omega_\text{n}^2$, we take only the $k=1$ mode (and so suppress index $k$). In this parametrization, the strength of the electron activated exchange terms, relative to driving the ESR resonantly at Rabi frequency $\Omega$, is conveniently captured by the dimensionless parameters $\eta^{\pm}_{j,k}$:

\begin{align*}
\begin{split}
  \eta^{\pm}_{\text{As},1} = &\frac{A_\text{nc,As,1}}{\omega_\text{n,As}} = \frac{A_\text{As}B_\text{Q,As}}{\omega_\text{n,As}^2} \sqrt{6N_\text{As}} \sin(2\theta_\text{As})\\
  \eta^{\pm}_{\text{As},2} = &\frac{A_\text{nc,As,2}}{2\omega_\text{n,As}} =\frac{A_\text{As}B_\text{Q,As}}{2\omega_\text{n,As}^2} \sqrt{6N_\text{As}} \cos^2(\theta_\text{As})\\
   \eta^{\pm}_{\text{In},1} = &\frac{A_\text{nc,In,1}}{\omega_\text{n,In}} =\frac{A_\text{In}B_\text{Q,In}}{\omega_\text{n,In}^2} \sqrt{\frac{1584N_\text{In}}{5}} \sin(2\theta_\text{In})\\
  \eta^{\pm}_{\text{In},2} = &\frac{A_\text{nc,In,2}}{2\omega_\text{n,In}} =\frac{A_\text{In}B_\text{Q,In}}{2\omega_\text{n,In}^2} \sqrt{\frac{1584N_\text{In}}{5}} \cos^2(\theta_\text{In})\text{.}
\end{split}
\end{align*}

\subsection{Master equation}
\label{subsec_mastereq}

We calculate the evolution of the system according to the following master equation:

\begin{align*}
\label{eq:master-eq}
\begin{split}
  \pdv{\rho(t|\delta)}{t}&=i[\rho(t|\delta),H_\text{tot}(\delta)]\\
  &+\sum_{j\in(\text{As},\text{In})}\sum_{l=-2}^{2} L(\sqrt{\frac{\Gamma_{\text{n},j}}{2}}\dyad{I_{z,j}+l})\\
  &+L(\sqrt{\frac{\Gamma_e}{2}}\sigma_z)\\
  &+L(\sqrt{\frac{\Omega}{2Q}}\sigma_x)\\
  &+L(\sqrt{\frac{\Gamma_\text{p}}{2}}\sigma_\text{+})
  \end{split}
\end{align*}

\noindent
where $\rho(t|\delta)$ is the density operator conditional on an initial detuning caused by the Overhauser field $\delta = 2 (A_\text{As} I_{z,\text{As}} + A_\text{In} I_{z,\text{In}})$; $L(a)=a\rho(t|\delta)a^\dagger -\{a^\dagger a,\rho(t|\delta)\}/2$ is the Lindblad operator; $\Gamma_\text{n}$ is broadening of the collective nuclear states,which acts as a pure dephasing term on the nuclei; $\Gamma_e$ is a pure dephasing rate on the electron ($L(\sigma_z)$). Importantly, the sum effect of electron and nuclear dephasing terms are constrained in simulation to replicate the experimental measurement via a Hahn Echo of $T_2 = 2000$\,ns, the electron's homogeneous dephasing time. $\Omega/Q$ represents a non-resonant, optical power-dependent electron spin relaxation mechanism ($L(\sigma_x)$), where $Q=Q(\Delta)$ is a proportionality factor between the relaxation rate and our optically-driven ESR that will depend on the Raman laser detuning $\Delta$, as identified in our previous work\cite{Bodey2019}; and $\Gamma_\text{p}$ is an optical electron spin pumping term ($L(\sigma_\text{+})$) that is turned on when initialising the electron and cooling the nuclear ensemble.


\subsection{Simulation}
\label{subsec_simulation}

\begin{figure*}
\includegraphics[width = 0.88\textwidth]{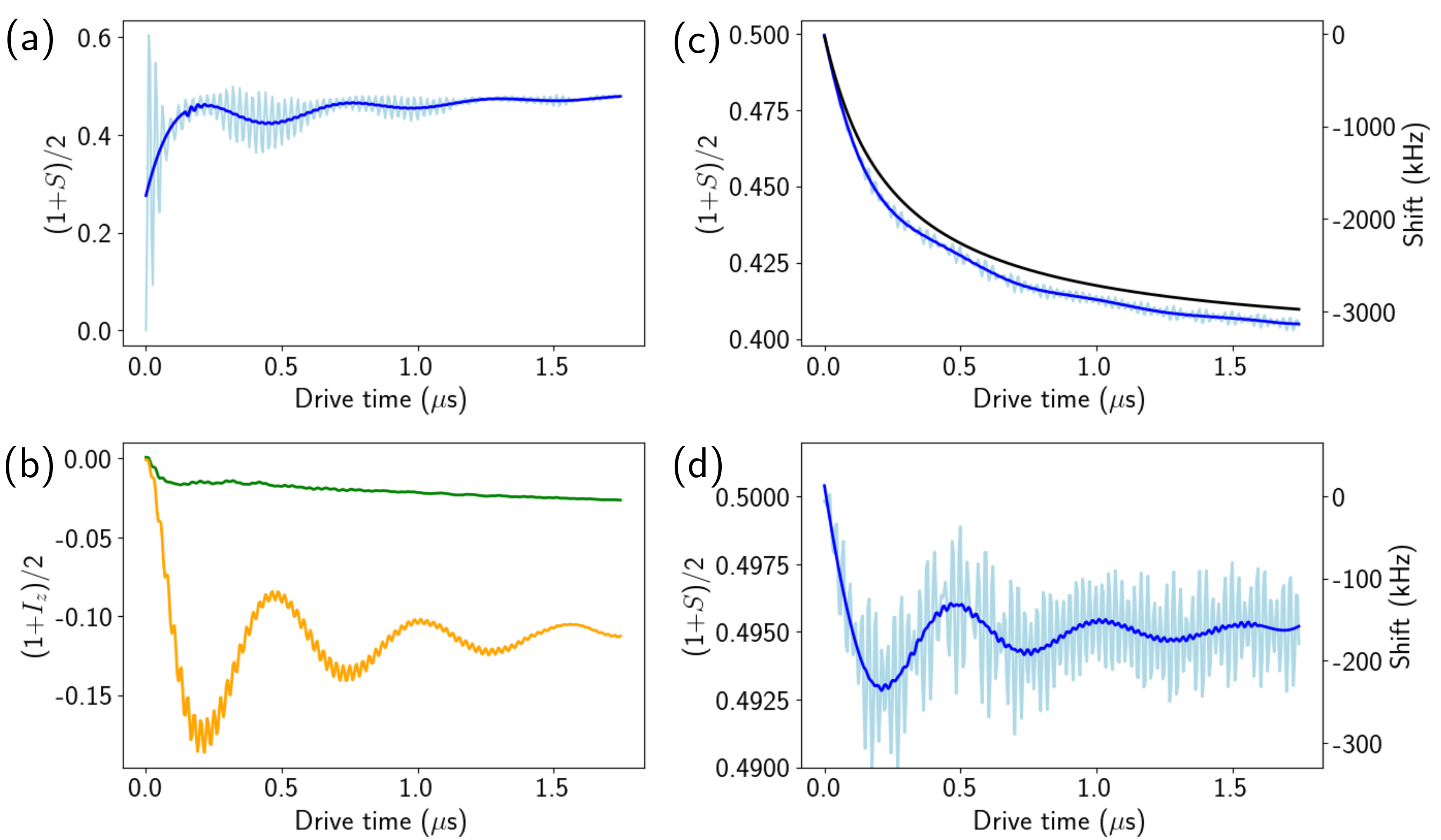}
\caption{\textbf{Simulation results.} \textbf{(a)} electron polarization directly following magnon injection (light blue). The dark blue curve is the same passed through a third-order Savitzky-Golay filter with a $300$\,ns window. \textbf{(b)} indium (orange) and arsenic (green) nuclear polarization directly following magnon injection \textbf{(c)} electron polarization following side-of-fringe Ramsey, \emph{reference} before magnon injection (black) and \emph{sense} following magnon injection (light blue). The dark blue curve is the \emph{sense} passed through a third-order Savitzky-Golay filter with a $300$\,ns window. \textbf{(d)} difference between the \emph{sense} and \emph{reference} curves from \textbf{c}. The dark blue curve is the same passed through a third-order Savitzky-Golay filter with a $300$\,ns window. Simulation parameters for all curves: $\Omega = 34$\,MHz, $\delta = 24$\,MHz, $\omega_\text{n,In} = 42$\,MHz, $\omega_\text{n,As}=32.5$\,MHz, $N_\text{In}=364$, $N_\text{As}=342$, $A_\text{In} = 0.4$\,MHz, $A_\text{As} = 1.35$\,MHz, $B_{Q,\text{In}} = 1$\,MHz, $B_{Q,\text{As}} = 1$\,MHz, $\theta_\text{In}=25^\circ$, $\theta_\text{As}=25^\circ$, $T_2^*=32$\,ns, $T_2=2000$\,ns, $Q=45$, $\Gamma_{\text{n,As}} = 0.05$\,MHz, $\Gamma_{\text{n,In}}=0.06$\,MHz.}
\label{simfig}
\end{figure*}

We use the QuTip\cite{Johansson2013} package for numerical integration of our quantum master equation. Each step of our pulse sequence is simulated using the master equation solver with the appropriately tuned Hamiltonian and dephasing terms.

To make the calculation tractable, we truncate the Hilbert space of the problem to capture only first-order transitions $M^\pm_{j,k}$ ($k\in1,2$). We calculate the dynamics for each species, $j$, only over the corresponding five-dimensional subspace $\{\ket{I_{z,j}+l}:l=-2,-1,0,+1,+2\}$, as shown in Fig.~\ref{fig_level_scheme}. Over this restricted space, we take the initial nuclear density operator to be a pure state $\ket{\uparrow,I_{z,\text{As}},I_{z,\text{In}}}$ (where the Overhauser shift $A_\text{As} I_{z,\text{As}} + A_\text{In} I_{z,\text{In}}$ makes the drive detuning $\delta=0$). We capture the effect of Overhauser field fluctuations -- leading to inhomogeneous dephasing of the electron spin ($T_2^*$) -- by taking realizations of the above master equation for values taken across the cooled Overhauser distribution 
\begin{equation*}
p(\delta) = \frac{1}{\sqrt{2\pi\sigma_\delta^2}}\exp(-\frac{\delta^2}{2\sigma_\delta^2})\text{,}
\end{equation*}
where $\sigma_\delta = \sqrt{2}/T_2^*$. The ensemble density operator, $\chi$, is then calculated by averaging the conditional density operator $\rho(t|\delta)$ over the different initial configurations $\chi(t)=\int\dd{\delta}p(\delta)\rho(t|\delta)$.

Figure \ref{simfig} presents the full results for a simulation where the drive time is varied and the system variables are extracted directly following magnon injection and following a side-of-fringe Ramsey sequence, as executed experimentally in the context of Fig. 4 of the main text. Here $\Omega = 34$\,MHz and $\delta = 21$\,MHz, which targets the Hartmann-Hahn resonance condition with the indium first sideband at $\omega_\text{In} = 42$\,MHz, as reported via the differential shift in Fig. 4b of the main text. 

\begin{figure*}
\centering
\includegraphics[width = 0.9\textwidth]{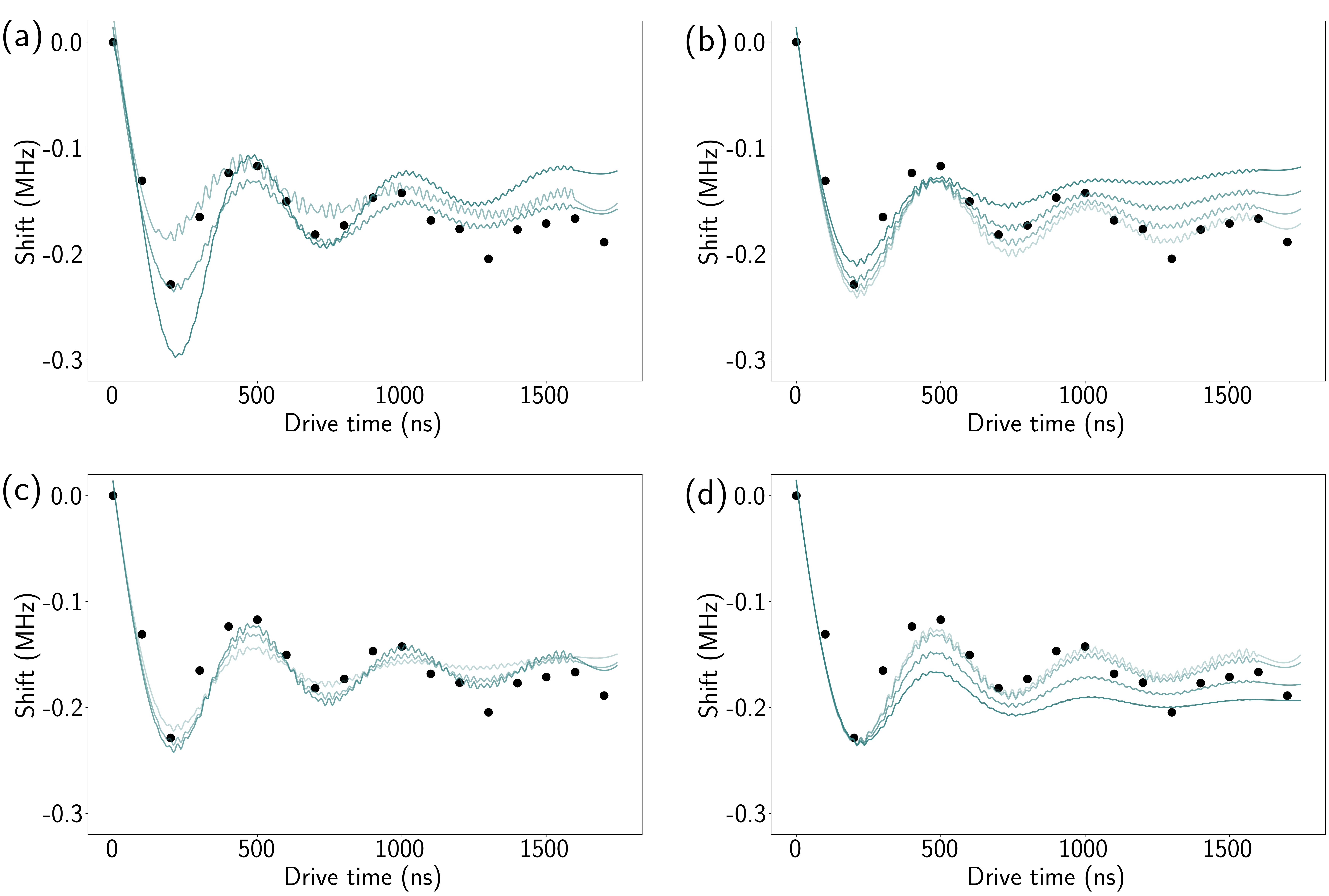}
\caption{\textbf{Effects of dephasing in simulations.} Black data points in all data are those reported in Fig. 4b of the main text ($\Omega=34$\,MHz, $\delta=21$\,MHz). Simulation parameters for all curves, unless otherwise stated: $\Omega = 34$\,MHz, $\delta = 24$\,MHz, $\omega_\text{n,In} = 42$\,MHz, $\omega_\text{n,As}=32.5$\,MHz, $N_\text{In}=364$, $N_\text{As}=342$, $A_\text{In} = 0.4$\,MHz, $A_\text{As} = 1.35$\,MHz, $B_{Q,\text{In}} = 1$\,MHz, $B_{Q,\text{As}} = 1$\,MHz, $\theta_\text{In}=25^\circ$, $\theta_\text{As}=25^\circ$, $T_2^*=32$\,ns, $T_2=2000$\,ns, $Q=45$, $\Gamma_{\text{n,As}} = 0.05$\,MHz, $\Gamma_{\text{n,In}}=0.06$\,MHz. All simulations are passed through a third-order Savitzky-Golay filter with a $300$\,ns window. \textbf{(a)} from light to dark: $T_2^*=15$\,ns, $T_2^*=30$\,ns, $T_2^*=60$\,ns. \textbf{(b)} from light to dark: $\Gamma_e=0.1$\,s$^{-1}$, $\Gamma_e=0.45$\,s$^{-1}$, $\Gamma_e=1$\,s$^{-1}$, $\Gamma_e=2$\,s$^{-1}$. \textbf{(c)} from light to dark: $Q=20$, $Q=45$, $Q=100$. \textbf{(d)} from light to dark: $\Gamma_n=0.001$\,MHz, $\Gamma_n=0.05$\,MHz, $\Gamma_n=0.25$\,MHz, $\Gamma_n=0.5$\,MHz.}
\label{dephasing}
\end{figure*}

\subsubsection{Coherence times}

Our dephasing parameters are verified against two independently measured quantities: the inhomogeneous dephasing time $T_2^*$, which we obtain by simulating a Ramsey experiment; the homogeneous dephasing time $T_2$, which we obtain by simulating a Hahn Echo. In both cases the simulation parameters are constrained such that these values agree with experimentally measured parameters.

\subsubsection{Spectrum}

We obtain our simulated spectrum by implementing the following pulse sequence: \textit{Reference} Ramsey sequence, electron re-initialization, magnon sideband drive at a fixed drive time and Rabi frequency $\Omega$ (same as experiment) and a variable detuning $\delta$, electron re-initialization, \textit{sense} Ramsey sequence. At each step, and in particular after the \textit{sense} Ramsey sequence, we can extract the electron polarization $S=2\langle S_z \rangle$ from the average density matrix $\chi(t)$, and plot it versus drive detuning.

\subsubsection{Time dependence}

We obtain our simulated drive-time dependence by implementing the following pulse sequence:  \textit{Reference} Ramsey sequence, electron re-initialization, magnon sideband drive at a fixed detuning $\delta$ and Rabi frequency $\Omega$ (matching the experimental settings) and for a variable drive time $T$, electron re-initialization, \textit{sense} Ramsey sequence. At each step, and in particular after the \textit{sense} Ramsey sequence, we can extract the electron polarization $S=2\langle S_z \rangle$ from the average density matrix $\chi(t)$, and plot it versus drive time $T$.

\subsubsection{Effects of dynamic nuclear polarization}

We account in simulations for the effects of the MHz-scale dynamic nuclear polarization present during our experiments (as reported in Fig. 3 of the main text) by fitting our experimental \emph{reference} signal as a function of drive time. We then feed this into our simulation as a drive-time dependent detuning $\delta$. As an example, this is shown in Fig. \ref{simfig}c, where the simulated \emph{reference} signal (black curve) depends on drive time and reproduces the expected MHz-scale ESR shift.

\subsubsection{Effects of dephasing parameters}

In figure \ref{dephasing}, we explore the dependence of magnon dynamics on four dephasing parameters: $\Gamma_n$, $\Gamma_e$, $Q$, and $T_2^*$.

\subsection{Summary of model parameters}
\label{subsec_parameters}

In Table \ref{table1}, we list a number of parameters which are used as fixed inputs to the model, based on previous studies or inferred from independent measurements.

From previous measurements of QDs on this wafer \cite{Stockill2016}, we consider a QD composition of In$_{0.5}$Ga$_{0.5}$As. Total hyperfine interaction strengths for InGaAs are well-known and size-independent constants: $11.1$\,GHz for arsenic and $13.5$\,GHz for indium \cite{Urbaszek2013}. Having pinned the composition, we can then calculate the number of nuclei in the QD from a measurement of the electron spin decoherence time $T_{2,\text{pre}}^* = 1.7$\,ns (from previous work \cite{Gangloff2019} on the same QD); in this case it is measured without any prior preparation of the nuclear spins to assess the ``natural" width of the Overhauser field distribution \cite{Urbaszek2013,Stockill2016}. Using a Gaussian electronic wavefunction whose standard deviation is half of the QD radius, we find that the dot must contain $\sim$80,000 nuclei. Accordingly, the number of arsenic nuclei is $N_\text{As}^\text{tot} =$ 40,000 and the number of indium nuclei is $N_\text{In}^\text{tot} =$ 20,000. Also present are 20,000 gallium nuclei, which we ignore in the magnon dynamics due to their typically much weaker quadrupolar contributions\cite{Stockill2016}. We fix the quadrupolar energy to be $B_\text{Q}=1$\,MHz for both species, as consistent with previous measurements \cite{Hogele2012,Stockill2016,Gangloff2019,Bodey2019}; to fit our data, the strength of the noncollinear interactions is varied instead via the collective enhancement $\sim\sqrt{N}$. We use nuclear gyromagnetic ratios of $7.22$\,MHz/T for arsenic, and $9.33$\,MHz/T for indium, to obtain our Zeeman energy splittings at $4.5$\,T of $\omega_\text{n,As} = 32.5$\,MHz and $\omega_\text{n,In}=42$\,MHz. The Hahn echo $T_2 = 2000 \pm 100$\,ns of the electron spin was also measured in previous work \cite{Gangloff2019}. The $Q$ factor of Rabi oscillations, equivalent to optically induced electron spin relaxation ($T_1$) is measured directly from the spectrum data of Fig. 2c in the main text; for these data, the far-detuned electron spin polarization, $S \approx -0.6$ after $1800$\,ns of drive time at $\Omega = 8$\,MHz, sets a relaxation rate of $\Omega/Q = \Omega/45 \approx 180$\,kHz. The Rabi frequency of $\Omega_\text{cool}=13$\,MHz and optical pumping rate of $\Gamma_\text{p}=35$\,MHz during nuclear spin cooling are known from Rabi oscillation and optical saturation calibration measurements, respectively. Following this preparation we measure electron spin decoherence time $T_{2}^* \approx 30$\,ns in a Ramsey measurement, as presented in Fig. 1 of the main text.

\begin{table}
\small
\begin{center}
\caption{Summary of independently measured or fixed model parameters}
\begin{tabular}[t]{ | c || c | }
\hline
Indium concentration In$_x$Ga$_{1-x}$As, $x$ &  $0.5$ \\ \hline
Zeeman energy (at $4.5$\,T), $\omega_\text{n,As}$, arsenic & $32.5$\,MHz \\ \hline
Zeeman energy (at $4.5$\,T), $\omega_\text{n,In}$, indium & $42$\,MHz \\ \hline
Total hyperfine interaction, arsenic & $11.1$\,GHz \\ \hline
Total hyperfine interaction, indium & $13.5$\,GHz  \\ \hline
Quadrupolar constant, arsenic, $B_\text{Q,As}$ & 1\,MHz \\ \hline
Quadrupolar constant, indium, $B_\text{Q,In}$ & 1\,MHz \\ \hline
Electronic decoherence time, pre-cooling, $T_{2,\text{pre}}^*$ & $1.7$\,ns \\ \hline
Electronic decoherence time, post-cooling, $T_2^*$ & $32$\,ns \\ \hline
Electronic Hahn Echo time, $T_2$ & $2000$\,ns \\ \hline
Electronic Rabi Quality Factor, $Q$ &  $45$ \\ \hline
Optical pumping rate (cooling), $\Gamma_\text{p}$ & $35$\,MHz \\ \hline
Cooling Rabi frequency, $\Omega_\text{cool}$ & $13$\,MHz \\ \hline
Total number of nuclei, $N_\text{As}^\text{tot}$, arsenic &  40,000 \\ \hline
Total number of nuclei, $N_\text{In}^\text{tot}$, indium &  20,000 \\ \hline
\end{tabular}
\label{table1}
\end{center} 
\end{table}

\begin{table}
\small
\begin{center}
\caption{Summary of fitted model parameters, magnon spectrum, main text Fig. 2}
\begin{tabular}[t]{ | c || c | }
\hline
Magnon participation ratio, arsenic, $N_\text{As}/N_\text{As}^\text{tot}$ & 0.05 \\ \hline
Magnon participation ratio, indium, $N_\text{In}/N_\text{In}^\text{tot}$ & 0.029 \\ \hline
Hyperfine constant per arsenic nucleus, $A_\text{As}$ & $950$\,kHz \\ \hline
Hyperfine constant per indium nucleus, $A_\text{In}$ & $550$\,kHz \\ \hline
Quadrupolar angle, arsenic, $\theta_\text{As}$ & 25$^{\circ}$ \\ \hline
Quadrupolar angle, indium, $\theta_\text{In}$ & 25$^{\circ}$ \\ \hline
Nuclear dephasing, arsenic, $\Gamma_{\text{n,As}}$ & $0.5$\,MHz \\ \hline
Nuclear dephasing, indium, $\Gamma_{\text{n,In}}$ & $0.6$\,MHz \\ \hline
\end{tabular}
\label{table2}
\end{center} 
\end{table}

\begin{table}
\small
\begin{center}
\caption{Summary of fitted model parameters, magnon dynamics, main text Fig. 4}
\begin{tabular}[t]{ | c || c | }
\hline
Magnon participation ratio, arsenic, $N_\text{As}/N_\text{As}^\text{tot}$ & 0.0045 \\ \hline
Magnon participation ratio, indium, $N_\text{In}/N_\text{In}^\text{tot}$ & 0.006-0.018 \\ \hline
Hyperfine constant per arsenic nucleus, $A_\text{As}$ & $1350(200)$\,kHz \\ \hline
Hyperfine constant per indium nucleus, $A_\text{In}$ & $500(60)$\,kHz \\ \hline
Nuclear dephasing, arsenic, $\Gamma_{\text{n,As}}$ & $0.05$\,MHz \\ \hline
Nuclear dephasing, indium, $\Gamma_{\text{n,In}}$ & $0.06$\,MHz \\ \hline
\end{tabular}
\label{table3}
\end{center} 
\end{table}

In Tables \ref{table2} and \ref{table3}, we list a number of parameters which are used as free parameters in the model. Table \ref{table2} reports the parameter values used in fitting the magnon spectrum of Fig. 2 in the main text. Table \ref{table3} reports the parameter values used in fitting the magnon dynamics of Fig. 4 in the main text. 

The single-nucleus hyperfine constants we report in the main text, $A_\text{As}$ and $A_\text{In}$, are from the fits to the spectra of Fig. 2c (electron polarization) and Fig. 2e (differential shift); we also report these values in Table \ref{table2}. $A_\text{As}$ and $A_\text{In}$ can also be fitted independently from the Rabi oscillations of Fig. 4b (indium $I_z \rightarrow I_z-1$ transition) and Fig. 4c (arsenic $I_z \rightarrow I_z-2$ transition); we report these values in Table \ref{table3} with errors representing one standard deviation of uncertainty on the mean. As a result, the values of $A_\text{As}$ and $A_\text{In}$ obtained from the spectrum (Fig. 2) are found to agree statistically with values obtained from Rabi oscillations (Fig. 4) within two standard deviations for arsenic and one standard deviation for indium. It is noteworthy that as direct observations of the hyperfine shift, the observed Ramsey shifts of Fig. 2e, Fig. 4b, and Fig. 4c strongly constrain the fitted values of $A_\text{As}$ and $A_\text{In}$ in our model.

The quadrupolar angles $\theta_\text{As}$ and $\theta_\text{In}$ are most strongly constrained in the fit to the spectrum of Fig. 2c (electron polarization), since within a species the ratio of $I_z \pm 1$ to $I_z \pm 2$ processes is fixed to $2\sin(2\theta)/\cos^2(\theta)$. We find an angle of $25^\circ$ for both species, which sets this ratio to $1.87$ as evidenced in the magnon sidebands of Fig. 2c. We keep these values fixed when fitting the magnon dynamics of Fig. 4.


The magnon participation ratios $N_\text{As}/N_\text{As}^\text{tot}$ and $N_\text{In}/N_\text{In}^\text{tot}$ represent the fraction of nuclei partaking in the arsenic and indium magnon modes, respectively, and are defined as the fitted number of nuclei over the number of nuclei estimated from a measurement of $T_2^*$ (see Table \ref{table1}). This parameter, tuned individually for each species, is how the activated exchange frequency can be fitted. Making this parameter a function of drive $\Omega$ captures the drive-dependence of the activated exchange frequency, as described in the context of Fig. 4d of the main text and of section \ref{sec_inhom} of this document.

\subsection{Estimate of bipartite electron-nuclear entanglement}
\label{subsec_entanglement}

Having fitted our experimental data with the observables extracted from the average density matrix $\chi(t)$, it is also possible from this density matrix to extract any quantum correlations present in the simulated system. In particular, having measured and fitted coherent magnon oscillations (Fig. 4 of the main text), we are interested in the level of bipartite entanglement present between the electron and a single magnon mode, as could be used towards a quantum memory. To do so, we define a target Bell state of form,

\begin{equation*}
\ket{\psi_\phi} = \ket{\tilde{\uparrow},I_{z,\text{As}}} + \exp(i\phi)\ket{\tilde{\downarrow},I_{z,\text{As}}-1}\text{,}
\end{equation*}
\noindent
where the electron state is expressed in the dressed state basis,
\begin{align*}
\label{eq:master-eq}
\begin{split}
\ket{\tilde{\uparrow}} &= \cos(\theta/2) \ket{\uparrow} + \sin(\theta/2)  \ket{\downarrow}\\
 \ket{\tilde{\downarrow}} &= \cos(\theta/2) \ket{\downarrow} - \sin(\theta/2)  \ket{\uparrow}\text{,}
  \end{split}
\end{align*}

\noindent
where $\sin(\theta) = \Omega/\sqrt{\Omega^2 + \delta^2}$. We then get our estimate of bipartite entanglement from the overlap fidelity, 

\begin{equation*}
\mathcal{F} = \underset{\phi}{\max} \bra{\psi_\phi} \chi(t) \ket{\psi_\phi}\text{.}
\end{equation*}

\begin{figure}
\includegraphics[width=\columnwidth]{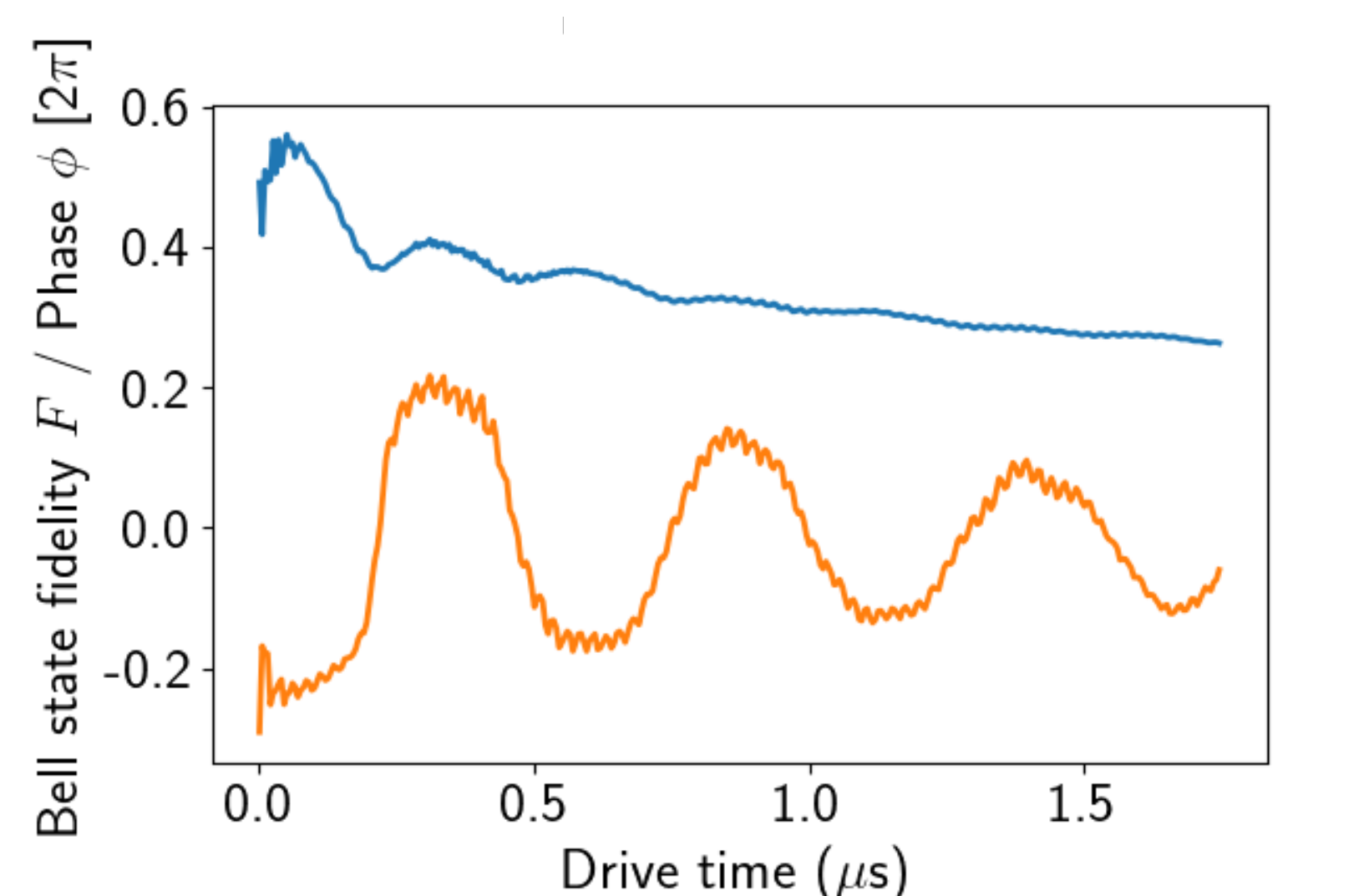}
\caption{\textbf{Bell state fidelity and phase.} Fidelity $\mathcal{F}$ (blue) and Bell state $\ket{\psi_\phi}$ phase $\phi$ (orange) as a function of drive time. Simulation parameters: $\Omega = 34$\,MHz, $\delta = 24$\,MHz, $\omega_\text{n,In} = 42$\,MHz, $\omega_\text{n,As}=32.5$\,MHz, $N_\text{In}=364$, $N_\text{As}=342$, $A_\text{In} = 0.4$\,MHz, $A_\text{As} = 1.35$\,MHz, $B_{Q,\text{In}} = 1$\,MHz, $B_{Q,\text{As}} = 1$\,MHz, $\theta_\text{In}=25^\circ$, $\theta_\text{As}=25^\circ$, $T_2^*=32$\,ns, $T_2=2000$\,ns, $Q=45$, $\Gamma_{\text{n,As}} = 0.05$\,MHz, $\Gamma_{\text{n,In}}=0.06$\,MHz.}
\label{entang}
\end{figure}

In Fig. \ref{entang}, we show the dependence of the fidelity $\mathcal{F}$ and state phase $\phi$ on drive time for the experiment performed in the context of Fig. 4b of the main text, whose full simulation results are shown in Fig. \ref{simfig}. The values shown in the inset of Fig. 4b of the main text are the maxima of fidelity across drive time.

\section{Collective Rabi oscillations with inhomogeneous broadening}
\label{sec_inhom}

\begin{figure*}
\centering
\includegraphics[width=\textwidth]{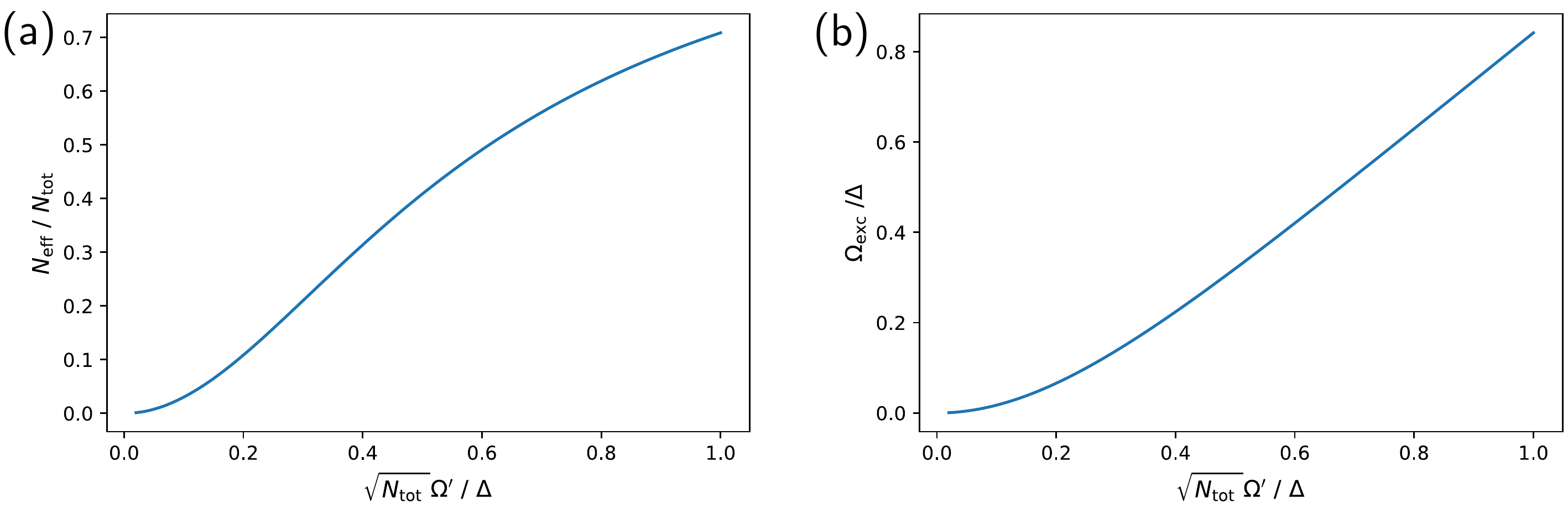}
\caption{\textbf{Collective Rabi oscillations with inhomogeneities.} \textbf{(a)} Effective number of nuclei in the magnon mode as a function of the normalized maximal collective Rabi frequency $\sqrt{N_\text{tot}}\Omega'/\Delta$. \textbf{(b)} Normalized electron-nuclear exchange frequency, $\Omega_\text{exc}/\Delta$, as a function of the normalized maximal collective Rabi frequency $\sqrt{N_\text{tot}}\Omega'/\Delta$.}
\label{fig_neff}
\end{figure*}

In the experiments, we measure a collective mode characterized by an electron-nuclear {\it activated exchange} frequency scaling as $\sqrt{N}\Omega'$, where $\Omega'$ is the single nucleus-electron Rabi coupling (tuned by the ESR drive strength $\Omega$). In the absence of inhomogeneities, all nuclei $N=N_\text{tot}$ take part identically, leading to a simple linear dependence of the {\it activated exchange} frequency on the ESR drive strength $\Omega$. However, in the presence of inhomogeneities, each nuclear spin couples to the ESR drive field differently due to its spectral detuning, $\delta$, as discussed already in Section \ref{sho_derivation}. However, the frequency scaling of the {\it activated exchange}  in the presence of inhomogeneities still stays linear, albeit with a reduced number of nuclei, $N_{\rm eff}<N_\text{tot}$. The superlinear scaling (with exponent value of $2.0$ for the $I_z-1$ mode and $1.1$ for the $I_z-2$ mode) observed in the experiments is not captured by the basic assumption of noninteracting nuclei. This points towards an $N_{\rm eff}$ value that is actually not constant, but is increasing with the ESR drive strength. In other words, Fig. 4d of the main text suggests that more nuclei take part coherently in the collective magnon mode as the ESR drive strength is increased.  

To mimic this behaviour, we introduce a transfer function whose width increases with the drive strength, $\Omega$. By assuming for the magnon mode a Lorentzian spectrum of width $\sqrt{N_{\mathrm{eff}}}\Omega'$, we can calculate the number of nuclear spins comprising the magnon mode self-consistently via
\begin{align}
 \label{SC}
 N_{\mathrm{eff}} = N_\text{tot}\int^{\infty}_{-\infty} \frac{N_{\mathrm{eff}}{\Omega'}^2}{N_{\mathrm{eff}}{\Omega'}^2+\delta ^2}S(\delta)d\delta \, , 
\end{align}
\noindent where $S(\delta)$ is the detuning probability distribution characterising the nuclear inhomogeneities. When we assume this distribution to be a Gaussian function, $S(\delta )=\frac{1}{\sqrt{\pi} \Delta}{\rm exp}(-\delta^2/\Delta^2)$, the self-consistent $N_{\rm eff}$ above is equivalent to
\begin{align}
 \label{SCgauss}
 \frac{N_{\mathrm{eff}}}{N_\text{tot}}=\frac{\sqrt{\pi N_{\mathrm{eff}}}\Omega'}{\Delta} {\rm exp}(\frac{N_{\mathrm{eff}} \Omega'^2}{\Delta^2}){\rm erfc}(\frac{\sqrt{N_{\mathrm{eff}}}\Omega'}{\Delta})
\end{align}
\noindent where ${\rm erfc}$ is the complementary error function. Solving this equation numerically, we find that the number of effective nuclei partaking in the collective mode $N_{\mathrm{eff}}$ increases with $\Omega'$, resulting in a superlinear increase of the collective Rabi frequency with the ESR drive strength (Fig. \ref{fig_neff}), qualitatively reproducing Fig. 4d of the main text. 

\noindent We emphasize that this behaviour of $N_{\rm eff}$ is qualitatively robust and does not depend critically on the specific choice of transfer function, as long as its width increases with drive strength, nor on the detuning probability distribution $S(\delta)$, chosen to be Gaussian here as an example. Such drive-dependent scaling of $N_{\mathrm{eff}}$ is a typical indication of synchronization (or entrainment) phenomenon \cite{Kuramoto2005}, which would indicate the presence of interactions among the nuclei. While the model in Section \ref{sho_derivation} of the SI does not include such interacting nuclei, a possible candidate to consider is the recently identified electron-mediated nuclear-nuclear interactions which scale as $A_j^2/\omega_\text{e}$ \cite{Wust2016}, analogous to the well-known RKKY interaction \cite{Ruderman1954}. Identifying the exact mechanism of nonlinearity in Fig. 4 of the main text will require further theoretical and experimental investigations, but the sensing capability demonstrated in this work promises more insights into this matter in a near future.


\bibliographystyle{Science.bst}
\bibliography{supplement}
\newpage